\documentclass[twocolumn]{article}
\usepackage{times}
\usepackage{graphicx}
\usepackage{url}
\usepackage{siunitx}
\usepackage{amsmath}
\usepackage{subcaption}
\usepackage{amsfonts}
\usepackage{gensymb}
\usepackage{braket}
\usepackage{makecell}
\usepackage{mwe}

\usepackage{float}
\usepackage{algorithm}
\usepackage{algorithmic}
\usepackage{array}

\usepackage{multirow}

\usepackage{booktabs}
\usepackage[numbers,sort&compress]{natbib}

\title{X-ray Emission Spectropolarimetry of Strongly Anisotropic Single Crystal Systems using a Rowland Circle Geometry}
\author{
  Jared E. Abramson$^{1,\ddag}$, 
  Charles A. Cardot$^{1,\ddag}$, 
  Josh J. Kas$^1$, 
  John J. Rehr$^1$,
  Werner Kaminsky$^2$,\\
  Herwig Michor$^3$,
  Marta Roman$^{4,5}$,
  Petra Becker$^5$,
  Gerald T. Seidler$^{1,*}$\\
  \\
  $^1$Department of Physics, University of Washington, Seattle, WA 98195, USA\\
  $^2$Department of Chemistry, University of Washington, Seattle, WA 98195, USA\\
  $^3$Institute of Solid State Physics, TU Wien, Wiedner Hauptstrasse 8-10, A-1040 Wien, Austria \\
  $^4$Institute of Physics and Applied Computer Science, Faculty of Applied Physics and \\ Mathematics, Gdansk University of Technology, Narutowicza 11/12, 80-233 Gdansk, Poland \\
  $^5$Institute of Geology and Mineralogy, University of Cologne, 50674 Cologne, Germany \\
  $^\ddag$These authors contributed equally to this work.\\
  $^*$seidler@uw.edu
}

\begin{document}

\twocolumn[
    \maketitle
    \begin{center}
        \begin{minipage}{0.8\textwidth} % Adjust the width of the abstract area
            \raggedright % Left-align the text
            Polarization dependence has historically seen extensive use in x-ray spectroscopy to determine magnetic and local geometric properties, but more broadly as a way to gain extra sensitivity to electronic structure at the level of individual magnetic orbitals. This is often done in the context of x-ray absorption through techniques like x-ray magnetic circular dichroism or x-ray linear dichroism, but it has seen little application to x-ray emission. Here we explore the information contained in the polarized emission of two 3$d$ transition metal systems across both core-to-core (CtC) and valence-to-core emission (VtC) lines. We demonstrate how the Rowland circle geometry can be used as a spectropolarimeter, and apply it to the x-ray emission spectroscopy of spin-1/2 Cu(II) and spin-0 Ni(II) ions in LiVCuO$_4$ and DyNiC$_2$, respectively. From this we explore how the polarized XES interrogates of the occupied density of states at the valence level, either as a second order effect through Coulomb exchange (CtC x-ray emission) or by direct transitions (VtC x-ray emission). We find that the polarized x-ray emission can provide insights into the valence electron orbital occupation, in much the same way that is achievable with polarized absorption or angle-resolved photoemission spectroscopy techniques. Finally, we highlight how the individually polarized dipole emission spectra can be extracted from a linearly independent suite of directed emission spectra, allowing for polarized measurements at high Bragg angle with lower experimental broadening.
        \end{minipage}
        \vspace{1em} % Adds space after the abstract
    \end{center}
]

\section{Introduction}

Many functional materials, such as superconducting cuprates, layered perovskites, and quasi-1D materials, exhibit anisotropic electronic and chemical environments \cite{evans2021layered, zhou-2022, schlappa2012spinorbital}. This makes them ideal candidates for polarization-sensitive x-ray techniques, which exploit dipole selection rules to resolve direction-dependent electronic features \cite{southworth1991anisotropy}. 
Polarized XAS, in particular, has become a standard technique in studying the magnetic, orbital, and local electronic anisotropy. It is the backbone to many polarized spectroscopy techniques such as x-ray magnetic circular dichroism (XMCD) \cite{stohr-1995, van-der-laan-2014}, x-ray magnetic linear dichroism (XMLD) \cite{heppell-2025}, angle-resolved photoemission spectroscopy (ARPES) \citep{sobota-2021, rosenberg-2022, bergner-2022, falub-2005, hajiri-2012}, and resonant inelastic x-ray scattering (RIXS) \cite{ament-2011, ulrich-2009, harada-2002, sala-2011, van-veenendaal-2006}.

X-ray emission spectroscopy (XES) is a versatile, element-specific probe that is well suited to a wide range of sample environments, including liquids, powders, single crystals, and even under \textit{in situ} or \textit{operando} conditions. The technique is ideal for characterizing the occupied density of states, providing insight into valence-level electronic structure, oxidation states, covalency, and chemical bonding \cite{glatzel-2004, lafuerza-2020}. Because XES exclusively probes occupied orbitals, it can be paired with x-ray absorption spectroscopy (XAS), which probes unoccupied orbitals, to create a powerful complementary framework for understanding electronic and chemical structure \cite{Kowalska-2015, krishnan-2024, giordanino-2014}.

In contrast, polarized non-resonant XES has seen little study due to the difficulty involved in measuring the polarization of a photon from low brilliance sources, i.e. fluorescence. Studies by Dr{\"a}ger and Czolbe \cite{drager-1984, czolbe-1992} showed there were polarization effects in XES which reflected the $m$-resolved density of states, but these developments were hampered due to optical constraints and detector efficiency. Later, Bergmann and co-workers \cite{bergmann-2002} demonstrated how electronic anisotropy from different ligand species can produce a polarization dependence in the ligand-coupled K$\beta''$ peak of Mn VtC-XES from [Rh(en)$_3$][Mn(N)(CN)$_5$]$\cdot$H$_2$O single crystals. 

To further investigate the application of polarized XES, we present two case studies on single crystal systems, LiVCuO$_4$ and DyNiC$_2$. LiVCuO$_4$ is a low-dimensional cuprate with strong orbital ordering from crystal field effects \cite{nishimoto-2012, huang-2011} and DyNiC$_2$ is an electronically quasi-one-dimensional rare-earth inter-metallic carbide \cite{roman-2023, maeda-2019}. Similar to other rare-earth nickel carbides such as SmNiC$_2$, the quasi-one-dimensionality of DyNiC$_2$ arises from its Ni chain structures, which drive anisotropic electronic behavior along the chain direction \cite{kim-2013}. We directly measure polarized XES using a spectropolarimeter design similar to the one developed by Dr{\"a}ger \textit{et al.} \cite{drager-1976} but with improved energy resolution, and indirectly by a new technique to extract the polarized spectra from a set of linearly independent, unpolarized XES spectra. We employ multiplet and real space Green's function calculations to help interpret the electronic structure information from polarized CtC- and VtC-XES techniques. From the lineshape and energy shifts of the polarized x-ray emission, we can infer the occupational configuration of the valence electrons to a similar extent as can be achieved with x-ray linear dichroism (XLD) or ARPES studies \cite{king-2014, wang-2023, huang-2004, wu-2004}.

% Finally, we propose future developments using asymmetric Rowland operation \cite{gironda-2024} to enable the resolution of more subtle polarization effects, including quadrupole contributions, thereby expanding the analytical power of polarized XES.

% Finally we propose future developments using asymmetric Rowland operation \cite{gironda-2024} to achieve better resolution and polarization sensitivity in spectropolarimetry, which would push the analytical capabilities of polarized XES into being a proper complement to polarized XAS.

\subsection{Manuscript Overview}

The manuscript proceeds as follows. Section \ref{sec:background} describes the relevant terminology, a framework of XES and corresponding toy model, and the single crystal systems we study. In Section \ref{sec:methods} we describe the experimental and computational methods. Special attention will be paid to the experimental setup and data processing to confirm that we perform a comparison of different polarizations on a consistent energy scale. In Section \ref{sec:res_and_disc} we present a polarization analysis of the CtC K$\beta$ XES and VtC-XES for both materials. Finally, in Section \ref{sec:conclusion} we summarize our results and conclude.

\section{Background}
\label{sec:background}

% \subsection{Terminology}
% \label{sec:terminology}

\subsection{X-ray Emission Spectroscopy}

X-ray emission is the fine energy resolution study of the fluorescence given off when an atom radiatively decays to fill a core hole left behind from an absorption event \cite{glatzel-2004}. Core-to-core (CtC) K$\beta$ XES involves filling a 1$s$ core hole from the 3$p$ orbital. CtC K$\beta$ XES has two main spectral features: K$\beta_{1,3}$ and K$\beta$'. The K$\beta_{1,3}$ is defined by transitions from the spin-orbit split 3$p_{1/2}$ or 3$p_{3/2}$ orbitals filling the 1$s$ hole. The K$\beta'$ satellite line originates from the exchange interaction between the 3$p$ core hole and unpaired 3$d$ electrons in the valence shell. The strength of the satellite is strongly dependent on the 3$d$ spin state, growing for higher spin systems \cite{glatzel-2004}. 

When the transition is from the valence levels, which have a mix of 3$d$ and ligand character, to 1$s$ core hole the process is known as a valence-to-core (VtC) transition. 3$d$ TMs VtC-XES also has two main spectral features; a main K$\beta_{2,5}$ spectral region coming from molecular orbitals with metal 3$d$, metal 4$p$, metal 4$s$, and ligand 2$p$ character, and a K$\beta''$ satellite from ligand 2$s$ electrons filling the metal 1$s$ hole. As such VtC-XES is highly depend on local environment with the K$\beta_{2,5}$ peak changing due to bonding and 3$d$ electron configuration and the K$\beta''$ peak energy and intensity being dependent on ligand speciation and bond length \cite{bergmann1999}.

\subsection{Polarization Dependence}

\subsubsection{Terminology}

Here we introduce and define the terminology for directed and polarized spectra. Spectra measured at high Bragg angle will contain approximately equal contributions from in-plane (p) and out-of-plane (s) polarizations, and are refered to as \textit{unpolarized}. Spectra measured at low Bragg angle will be dominated by the out-of-plane polarization, and we call these \textit{partially polarized}. Further details will be provided in Section \ref{sec:expr_setup}.

% A key note is that both of these are considered "directed" spectra, where the specified axis refers to the direction of propagation of the photon (ex $I_x$). However, the "partially polarized" spectra are dominated by a specific polarization which is perpendicular to the direction of propagation (ex: $\sigma_z$, $I_x$). The "polarized" label refer to the spectrum created from a dipole transition along a specific axis. All purely "polarized" spectra (ex: $\sigma_x$) presented are either extracted from experiment following Section \ref{sec:dipole_spectra_extraction} or calculated from theory.

A key note is that both of these are considered \textit{directed} spectra, where the specified axis refers to the direction of photon propagation, denoted as $I_x$, $I_y$, or $I_z$. However, the \textit{partially polarized} spectra are dominated by a specific polarization component that is perpendicular to the propagation direction, and are therefore denoted with both the polarization axis and propagation axis specified (ex: $\sigma_x$, $I_z$). In contrast, \textit{polarized} spectra refer to emission resulting from a dipole transition along a specific axis, denoted purely as $\sigma_x$, $\sigma_y$, or $\sigma_z$. All polarized spectra presented in this work are either extracted from experiment following Section \ref{sec:dipole_spectra_extraction} or directly calculated from theory. The terminology is summarized in Table \ref{tab:terminology}.

\begin{table}[h]
\centering
\begin{tabular}{ll}
\hline
\textbf{Term} & \textbf{Definition} \\
\hline
Unpolarized & High Bragg Angle \\
Partially Polarized & Low Bragg Angle \\
$\sigma_x$ & $x$-polarized spectrum \\
$\sigma_y$ & $y$-polarized spectrum \\
$\sigma_z$ & $z$-polarized spectrum \\
$I_x$ & $x$-directed spectrum \\
$I_y$ & $y$-directed spectrum \\
$I_z$ & $z$-directed spectrum \\
\hline
\end{tabular}
\caption{Terminology and symbol definitions used in this work.}
\label{tab:terminology}
\end{table}

\subsubsection{Polarized Dipole Components}
\label{sec:selection_rules}

The x-ray emission intensity  from a given initial state $i$ is expressed in Eq. \ref{eq:FermisGoldenRule}. The total spectrum comes from the sum over Fermi's golden rule for all final states $f$ \cite{de-groot-2008}. The energy $\hbar \omega$ is the energy of the radiation, $m$ is the mass of the electron, $\vec{r}$ is the spatial coordinate vector, and $\vec{k}$ ($|\vec{k}| = 2\pi/\lambda$) is the propagation vector of the photon \cite{laihia-1998, de-groot-2008, ogasawara-2004}. The orientation of the emitted radiation is entirely described by $\vec{k}$ and the polarization unit vector $\hat{\epsilon}$, which is orthogonal to $\vec{k}$.

\begin{equation}
\begin{aligned}
    \sigma &= \frac{\omega e^2}{h m^2 c^3} \sum_f \left| \bra{f} \exp(i \vec{k} \cdot \vec{r}) \; \hat{\epsilon} \cdot \vec{r} \ket{i} \right|^2 \\ 
    &\times \delta(E_f - E_i + \hbar \omega)
    \label{eq:FermisGoldenRule}
\end{aligned}
\end{equation}

% If we approximate the wavelength of the photon, $\lambda$, to be much larger than the size of the 1$s$ shell, $\lambda \gg r$, we can expand the exponential function as a power series, $\exp(i \vec{k} \cdot \vec{r}) \approx 1 + i \vec{k} \cdot \vec{r} + (\vec{k} \cdot \vec{r})^2/2 + \dots$. Keeping only the first term gives the dipole approximation as shown in Eq. \ref{eq:DipoleApprox}, where the dipole transition operator in the matrix element can be expanded into Cartesian coordinates, $\hat{\epsilon} \cdot \vec{r} = \epsilon_x \hat{x} + \epsilon_y \hat{y} + \epsilon_z \hat{z}$. The "hat" symbol here denotes an operator. The isotropic spectrum is equivalent to taking the average of the trace of the polarization tensor.

If we approximate the wavelength of the photon, $\lambda$, to be much larger than the size of the 1$s$ shell, $\lambda \gg r$, we arrive at the dipole approximation in Eq. \ref{eq:DipoleApprox}. The dipole transition operator is given by $\sum\limits_{\alpha} \hat{\epsilon}_\alpha \cdot \vec{r}$. The isotropic spectrum is equivalent to taking the average of the trace of the polarization tensor.

% \begin{equation}
% \begin{aligned}
%     \sigma^{\textbf{dip}} &\propto \sum_f | \bra{f}  \sum_\alpha \hat{\epsilon}_\alpha \cdot \vec{r} \ket{i} |^2 \delta(E_f - E_i + \hbar \omega) \\
%     &\propto \sum_f \left| \bra{f} \epsilon_x \hat{x} + \epsilon_y \hat{y} + \epsilon_z \hat{z} \ket{i} \right|^2 \delta(E_f - E_i + \hbar \omega) \\
%     &= \sigma^{\textbf{dip}}_x + \sigma^{\textbf{dip}}_y + \sigma^{\textbf{dip}}_z
% \end{aligned}
% \label{eq:DipoleApprox}
% \end{equation}

\begin{equation}
\begin{aligned}
    \sigma^{\textbf{dip}} &\propto \sum_f | \bra{f}  \sum_\alpha \hat{\epsilon}_\alpha \cdot \vec{r} \ket{i} |^2 \delta(E_f - E_i + \hbar \omega) \\
    &\propto \sum_f \left| \bra{f} \epsilon_x \hat{x} + \epsilon_y \hat{y} + \epsilon_z \hat{z} \ket{i} \right|^2 \delta(E_f - E_i + \hbar \omega) \\
    &= \sigma_x + \sigma_y + \sigma_z
\end{aligned}
\label{eq:DipoleApprox}
\end{equation}

We refer to the observables $\sigma_x$, $\sigma_y$, and $\sigma_z$ as the polarized spectra. They represent the emission intensity as a function of energy for a dipole transition along a given axis, and they are the quantity that polarization analysis of a single-crystal XES study seeks to determine. For information about how the radiation pattern of dipole (and higher order) transitions, please refer to Supplemental Information section \ref{app:POW_SPEC_APP}. The dipole selection rules restrict transitions to be between orbitals that are related by $\Delta l = \pm 1$, which follows from the fact that the position operators $x$, $y$, and $z$ transform as components of a spherical tensor of rank 1 and thus only connect states whose angular momenta differ by one unit \cite{sakurai-2020}.

The spectrum corresponding to a photon with wave vector $\vec{k}$ will have a polarization perpendicular to the direction of propagation  \cite{griffiths-2017}. The intensity of the emitted photon is symmetric azimuthally about the dipole transition moment axis, which means that the measured emission along a given axis is composed of an equal mixing of the two signals with polarization perpendicular to the direction of propagation. The system of equations which gives the unpolarized intensity measured from a photon propagating in a given direction ($I_x$, $I_y$, and $I_z$) is shown in Eq. \ref{eq:dipole_eq_sys}, where $\sigma_x$, $\sigma_y$, and $\sigma_z$ correspond to the spectra from the $x$, $y$, and $z$ dipole transition operators respectively.

%We refer to these $I_x$, $I_y$, and $I_z$ as the unpolarized spectra. The system of equations which gives the unpolarized intensity measured from a photon propagating in a given direction is shown in Eq. \ref{eq:dipole_eq_sys}, where $\sigma_x$, $\sigma_y$, and $\sigma_z$ correspond to the spectra from the $x$, $y$, and $z$ dipole transition operators respectively.

\begin{equation}
    \vec{I}_{\text{dir}} =
    \begin{bmatrix}
    I_{x} \\
    I_{y} \\
    I_{z}
    \end{bmatrix}
    \propto
    \frac{1}{2}
    \begin{bmatrix}
    0 & 1 & 1 \\
    1 & 0 & 1 \\
    1 & 1 & 0
    \end{bmatrix}
    \begin{bmatrix}
    \sigma_x \\
    \sigma_y \\
    \sigma_z
    \end{bmatrix}
    \label{eq:dipole_eq_sys}
\end{equation}

% The directional dependence of a dipole emission is only dependent on the polarization of the light $\vec{\epsilon}$, which will always be perpendicular to the direction of propagation $\vec{k}$ \cite{griffiths-2017}. The isotropic nature of dipole emission in the plane perpendicular to the oscillating dipole means that, for emission along a given axis, the spectrum is composed of an equal mixing of any two perpendicular dipole transitions (in the plane). For emission along the coordinate axes, this can be represented by the expressions in Eq. \ref{eq:dipole_eq_sys}. Given a complete set of orthogonal spectra, the individual polarizations can be extracted by taking linear combinations of the individual spectra (ex: $x_{pol} = (y_{dir} + z_{dir} - x_{dir}) / 2$).

% \begin{equation}
%     \fontsize{10}{10}\selectfont
%     x_{dir} = \frac{1}{\sqrt{2}} 
%     \begin{bmatrix}
%         0 \\ 1 \\ 1
%     \end{bmatrix}, \;
%     y_{dir} = \frac{1}{\sqrt{2}} 
%     \begin{bmatrix}
%         1 \\ 0 \\ 1
%     \end{bmatrix}, \;
%     z_{dir} = \frac{1}{\sqrt{2}} 
%     \begin{bmatrix}
%         1 \\ 1 \\ 0
%     \end{bmatrix}
%     \label{eq:dipole_eq_sys}
% \end{equation}

\subsection{Toy Model of CtC-XES}

\label{sec:toymodel}

Local anisotropy is often reflected in the (projected) electronic density of states of the valence shell of metal ions. For the purpose of demonstrating the origin of polarization effects in CtC-XES, we investigate how anisotropy in the valence level electron configuration is transferred to other levels via the the electron-electron Coulomb interaction. This will be crucially instructive for demonstrating how the electron configuration can be inferred from the polarized CtC-XES of real systems, as we will see in section \ref{sec:res_and_disc}.

We consider a toy system of one $d$ electron and one $p$ electron, fixing the $d$ electron to be spin down in the $d_{xy}$ orbital. The choice of $d_{xy}$ for this example is arbitrary, but in a real system will be determined by valence level splitting from crystal field effects. The Coulomb Hamiltonian is given in second quantization in Eq \ref{eq:toy_model_2ndquant}, where $\tau = \sigma, m, l, n$ denotes the spin and orbital degrees of freedom. The two particle Coulomb operator between the $p$ and $d$ orbitals can be split into spherical ($\Theta^k[\tau_1 \tau_2 \tau_3 \tau_4]$) and radial ($R^k[\tau_1 \tau_2 \tau_3 \tau_4]$) components (Eq. \ref{eq:toy_model_expansion}), and further split into the `direct' ($H^{(0)}_{Fpd}$, $H^{(2)}_{Fpd}$) and `exchange' ($H^{(1)}_{Gpd}$, $H^{(3)}_{Gpd}$) terms (Eq. \ref{eq:toy_model_pdsimplify}) \cite{slater-1960}. The Slater-Condon $F$ and $G$ terms are usually determined by the overlap of the radial wave functions, but in this toy Hamiltonian we just set $R^{k} = [F^0_{pd}, G^1_{pd}, F^2_{pd}, G^3_{pd}]=[1.2, 8.0, 9.0, 5.0]$ to approximate the interaction between the 3$p$ and 3$d$ orbitals in the final state of a K$\beta$ XES process.

% The coefficients $\kappa_i$ set the scale of the individual Slater-Condon terms. For the toy Hamiltonian, we set the coefficients $\vec{\kappa} = [1.2, 8.0, 8.0, 5.0]$ to approximate the interaction between the 3$p$ and 3$d$ orbitals in the final state of a K$\beta$ XES process.

% \begin{subequations}
% \label{eq:toy_atomic}
% \begin{align}
%     &H_{\textrm{toy}}^{C} = \sum_{\tau_1 \tau_2 \tau_3 \tau_4} U_{\tau_1 \tau_2 \tau_3 \tau_4} \; a_{\tau_1}^{\dagger} a_{\tau_2}^{\dagger} a_{\tau_4} a_{\tau_3}
% \end{align}
% \begin{align}
%     &U_{\tau_1 \tau_2 \tau_3 \tau_4} = -\frac{1}{2} \delta_{\sigma_1, \sigma_3} \delta_{\sigma_2, \sigma_4} \sum_{k=0}^{\infty} \frac{4 \pi}{2k + 1} \\
%     &\times \bra{Y_{m_1}^{(l_1)}} Y_{m_1 - m_3}^{(k)} \ket{Y_{m_3}^{(l_3)}}
%     \bra{Y_{m_4}^{(l_4)}} Y_{m_4 - m_2}^{(k)} \ket{Y_{m_2}^{(l_2)}} \notag \\ &\times R^k[\tau_1 \tau_2 \tau_3 \tau_4] \notag
% \end{align}
% \begin{align}
%     U_{\tau_1 \tau_2 \tau_3 \tau_4} &= \kappa_{0} \, \textrm{F}0_\textrm{pd} + \kappa_{1} \, \textrm{F}2_\textrm{pd} \\
%     &+ \kappa_{2} \, \textrm{G}1_\textrm{pd} + \kappa_{3} \, \textrm{G}3_\textrm{pd} \notag
% \end{align}
% \end{subequations}

\begin{subequations}
\label{eq:toy_atomic}
\begin{align}
    \label{eq:toy_model_2ndquant}
    &H_{\textrm{toy}}^{C} = \sum_{\tau_1 \tau_2 \tau_3 \tau_4} U_{\tau_1 \tau_2 \tau_3 \tau_4} \; a_{\tau_1}^{\dagger} a_{\tau_2}^{\dagger} a_{\tau_4} a_{\tau_3}
\end{align}
\begin{align}
    \label{eq:toy_model_expansion}
    U_{\tau_1 \tau_2 \tau_3 \tau_4} &= -\frac{1}{2} \delta_{\sigma_1, \sigma_3} \delta_{\sigma_2, \sigma_4} \sum_{k=0}^{\infty} \frac{4 \pi}{2k + 1} \\
    &\times \bra{Y_{m_1}^{(l_1)}} Y_{m_1 - m_3}^{(k)} \ket{Y_{m_3}^{(l_3)}}
    \bra{Y_{m_4}^{(l_4)}} Y_{m_4 - m_2}^{(k)} \ket{Y_{m_2}^{(l_2)}} \notag \\ &\times R^k[\tau_1 \tau_2 \tau_3 \tau_4] \notag \\
    &= -\frac{1}{2} \delta_{\sigma_1, \sigma_3} \delta_{\sigma_2, \sigma_4} \sum_{k=0}^{\infty} \frac{4 \pi}{2k + 1} \notag \\
    &\times \Theta^k[\tau_1 \tau_2 \tau_3 \tau_4] \times R^k[\tau_1 \tau_2 \tau_3 \tau_4] \notag
\end{align}
\begin{align}
    \label{eq:toy_model_pdsimplify}
    H_{\textrm{toy}}^{C} &= -2\pi \sum_{\tau_1 \tau_2 \tau_3 \tau_4} \Bigg( 
        \Theta^0 F^0_{pd} 
        + \frac{1}{3} \Theta^1 G^1_{pd} \\
    &+ \frac{1}{5} \Theta^2 F^2_{pd} 
        + \frac{1}{7} \Theta^3 G^3_{pd} \Bigg) \; a_{\tau_1}^{\dagger} a_{\tau_2}^{\dagger} a_{\tau_4} a_{\tau_3} \notag \\
    &= H_{Fpd}^{(0)} + H_{Gpd}^{(1)} + H_{Fpd}^{(2)} + H_{Gpd}^{(3)} \notag
\end{align}
\end{subequations}

\begin{figure}
    \centering
    \includegraphics[width=1\linewidth]{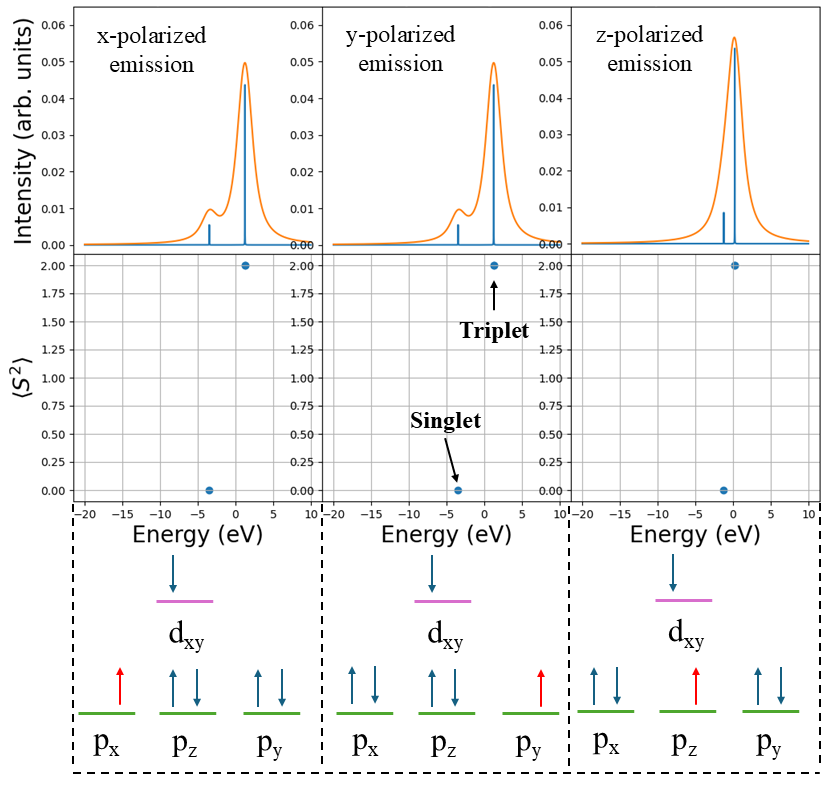}
    \caption{The polarized dipole $p \rightarrow s$ emission spectra and final states of our toy system described by Eq. \ref{eq:toy_atomic}. Only final state configurations with a spin down electron in the $d_{xy}$ orbital are calculated. The top row shows the spectra for different dipole transition operators ($x$, $y$, and $z$). The middle row shows the expectation value of the total system $\braket{S^2}$ operator for the final states. The last row shows the singlet configuration corresponding to the $p$ core hole created by each of the dipole transition operators}
    \label{fig:Kbeta_toy}
\end{figure}

\begin{figure}
    \centering
    \includegraphics[width=1\linewidth]{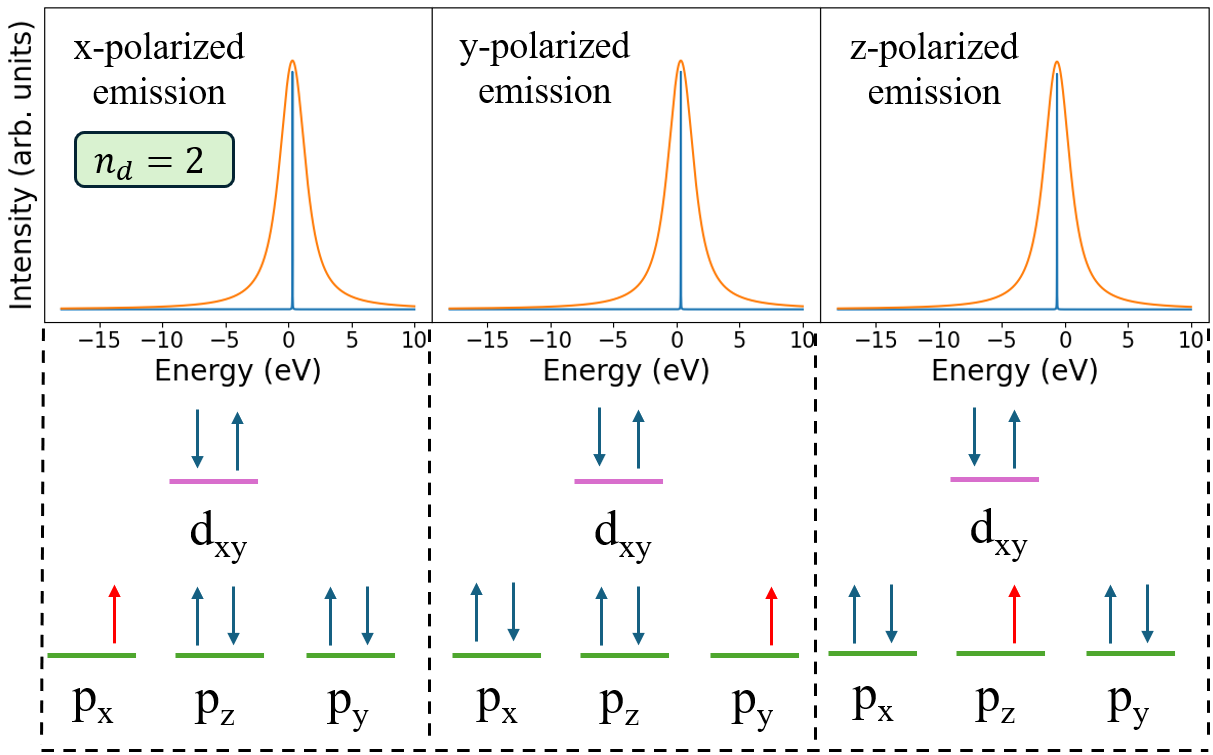}
    \caption{The polarized dipole $p \rightarrow s$ emission spectra for a final state described by Eq. \ref{eq:toy_atomic}, but with two electrons pinned into the same $d_{xy}$ orbital. The top row shows the emission spectra for each polarization and the bottom row shows the singlet configuration of each final state.}
    \label{fig:Twoelec_toy_model}
\end{figure}

\begin{table*}[ht]
\centering
\renewcommand{\arraystretch}{1.2} % Increases row height slightly
\begin{tabular}{|c|c|c|c|c|c|c|c|c|c|}
\hline
\textbf{$d$-Orbital} & \textbf{$d$-Spin} & \textbf{$p$-Orbital} & \textbf{$p$-Spin} & \makecell{\textbf{Spin} \\ \textbf{Aligned?}} & \textbf{$\braket{H_{\textrm{toy}}^{C}}$} & $\braket{H_{Fpd}^{(0)}}$ & $\braket{H_{Fpd}^{(2)}}$ & $\braket{H_{Gpd}^{(1)}}$ & $\braket{H_{Gpd}^{(3)}}$ \\ \hline
$d_{xy}$ & down & $p_x$ & down & yes & -0.885 & 0.748 & 0.457 & -1.600 & -0.490 \\ \hline
$d_{xy}$ & down & $p_x$ & up & no & 1.205 & 0.748 & 0.457 & 0.000 & 0.000 \\ \hline
$d_{xy}$ & down & $p_y$ & down & yes & -0.885 & 0.748 & 0.457 & -1.600 & -0.490 \\ \hline
$d_{xy}$ & down & $p_y$ & up & no & 1.205 & 0.748 & 0.457 & 0.000 & 0.000 \\ \hline
$d_{xy}$ & down & $p_z$ & down & yes & -0.473 & 0.748 & -0.914 & 0.000 & -0.306 \\ \hline
$d_{xy}$ & down & $p_z$ & up & no & -0.167 & 0.748 & -0.914 & 0.000 & 0.000 \\ \hline
\end{tabular}
\caption{Table of Coulombic $d$-orbital and $p$-orbital interactions for a system with one $d$ electron constrained in the spin down $d_{xy}$ fermionic mode and one $p$ electron.}
\label{tab:orbital_interactions}
\end{table*}

Table \ref{tab:orbital_interactions} shows the contributions to the total energy \textbf{$\braket{H_{\textrm{toy}}^{C}}$} for different $p$, $d$ configurations. These were calculated using the many-body second quantization code Quanty to encode the toy Hamiltonian \cite{haverkort-2012}. The direct terms $H^{(0)}_{Fpd}$ and $H^{(2)}_{Fpd}$ are always non-zero, with the $H^{(0)}_{Fpd}$ term corresponding to a spherically symmetric constant contribution and the $H^{(2)}_{Fpd}$ changing signs and magnitude according to whether the occupied orbitals have overlapping symmetry. This behavior is effectively an inter-orbital Hund's rule \cite{sajeev-2008}; when the $d$ and $p$ electrons have overlapping symmetry (ex: $d_{xy}$ and $p_x$) the energy of that configuration is raised compared to when their symmetries do not overlap (ex: $d_{xy}$ and $p_z$).%This behavior is effectively an inter-orbital Hund's rule \cite{sajeev-2008}, where when the $d$ and $p$ electrons have overlapping symmetry (ex: $d_{xy}$ and $p_x$). This raises the energy of that configuration versus when their symmetries do not overlap (ex: $d_{xy}$ and $p_z$). 

The exchange terms $H^{(1)}_{Gpd}$ and $H^{(3)}_{Gpd}$ come from the fermionic behavior of electrons which energetically split spin-aligned configurations from spin-opposed. Hund's principle of maximum multiplicity leads to systems with aligned spins having lower energy than when spins are opposed. The exchange terms behave similarly to the direct terms in that the magnitude of the interaction is larger when the symmetries of the two occupied orbitals are overlapping, but the contribution to the total energy is always negative. The behavior of the exchange Coulomb terms is what gives sensitivity to spin in CtC-XES, where for example the K$\beta'$ peak changes depending on whether a system is in a high spin or low spin configuration \cite{glatzel-2004}. 

We demonstrate how this behavior in Fig. \ref{fig:Kbeta_toy} combines with the difference in Coulomb interaction based on symmetry overlap to give polarized spectra. Starting from an initial state with six $p$ electrons and an $s$ core hole, we use the $x$, $y$, and $z$ dipole transition operators to control where the unpaired electron in the $p$ orbital ends up in the final state. The final state is constrained to have the single $d$ electron in the $d_{xy}$ orbital, meaning that our final states are analogous to the six rows shown in Table \ref{tab:orbital_interactions} (up to an overall shift). Note that the peak corresponding to a spin-aligned (triplet) final state is higher energy then the spin-opposed (singlet) peak. This is because the emission energy is equal to the difference in energies between the initial and final states.

Quanty was used to calculate the $p \rightarrow s$ polarized dipole emission spectra. The first row in Fig. \ref{fig:Kbeta_toy} shows $x$, $y$, and $z$ polarized emission spectra. We note that the $x$ and $y$ polarized emission are identical with a large splitting between the singlet and triplet states, while the $z$ polarized emission has a much smaller splitting. An important note is that because we are neglecting spin-orbit splitting in this toy model the multiplicity of the entire system $\braket{S^2}$ is a good quantum number, and it allows us to distinguish the configurations as seen in the second row. The singlet configuration for each final state is shown in the last row. For more information about the quantitative behavior of the splitting between the singlet and triplet states we refer readers to the theory discussion in Lafuerza \textit{et al.} \cite{lafuerza-2020}. The splitting between these configurations is the same mechanism that underpins the spin-dependence of the K$\beta$' peak in 3$p$ $\rightarrow$ 1$s$ XES. The behavior of this toy system highlights a key result, namely that the intensity and position of the K$\beta$' in polarized emission can be a reflection of orbital occupation as well as the spin state.

A simpler, but equally useful result comes from adding a second electron into the $d_{xy}$ orbital, see Fig. \ref{fig:Twoelec_toy_model}. The addition of a second electron takes the system from a spin-1/2 to spin-0 configuration and removes the split singlet-triplet behavior. While the final states of the non-degenerate polarizations will still experience an overall shift in energy, the effective anisotropy in the spectra becomes much weaker. 

The toy model here is simple, but the results are generic. We have demonstrated how the polarized emission of CtC-XES is sensitive to the occupation of the valence orbitals via the Coulomb exchange interaction. Consequently, any symmetry breaking perturbations (crystal field, charge transfer) will also be reflected in the spectra. This is distinct from the polarization sensitivity observed in absorption techniques like L$_{2,3}$ XLD or XMCD, where the electron transitions from $p$ into $d$ provide a direct probe of the unoccupied density of states in the presence of a core hole.

\subsection{Crystal Systems}
\label{sec:crystalsystems}

\begin{figure}
    \centering
    \includegraphics[width=1\linewidth]{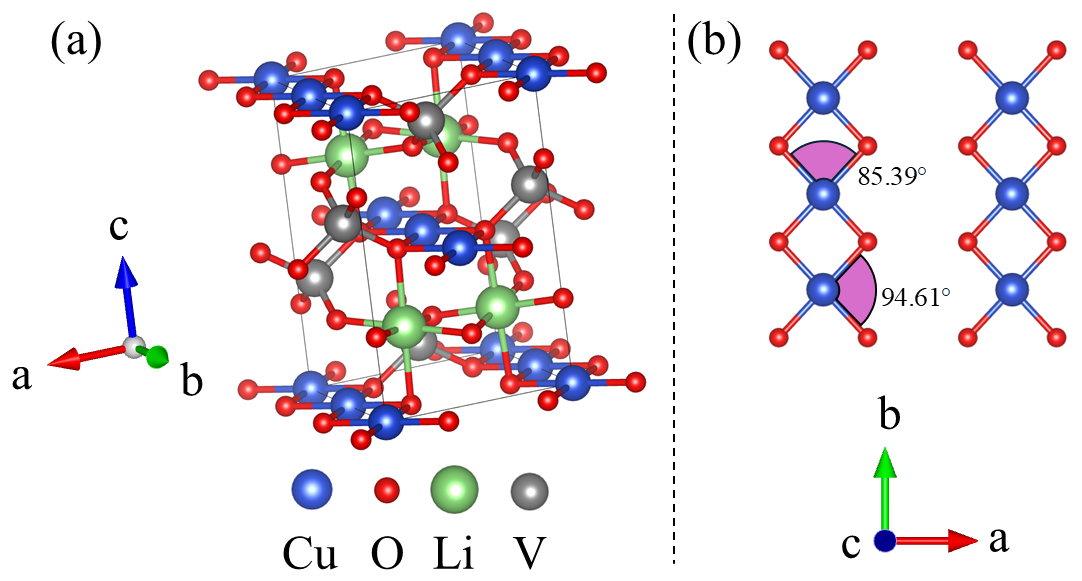}
    \caption{(a) LiVCuO$_4$ crystal structure. (b) Face down view on the a-b plane of square planar Cu-O chains.}
    \label{fig:LiVCuO4_crystal_structure}
\end{figure}

We study two transition metal compounds: LiVCuO$_4$ and DyNiC$_2$. Both materials are studied at room temperature. The crystal structure of LiVCuO$_4$ \cite{nishimoto-2012} is shown in Fig. \ref{fig:LiVCuO4_crystal_structure}. CuO$_4$ ladders run through the $ab$ plane, and have a nearly perfect square planar bond orientation with a small rhomboidal distortion, giving the Cu-O cluster a point group of D$_{2h}$. The Cu(II) ion leaves the system with $n_d = 9$. The crystal field splitting for the local cluster around the Cu leaves a single unpaired $d$-electron and a spin-1/2 system \cite{watanabe-1966}.

\begin{figure}
    \centering
    \includegraphics[width=1\linewidth]{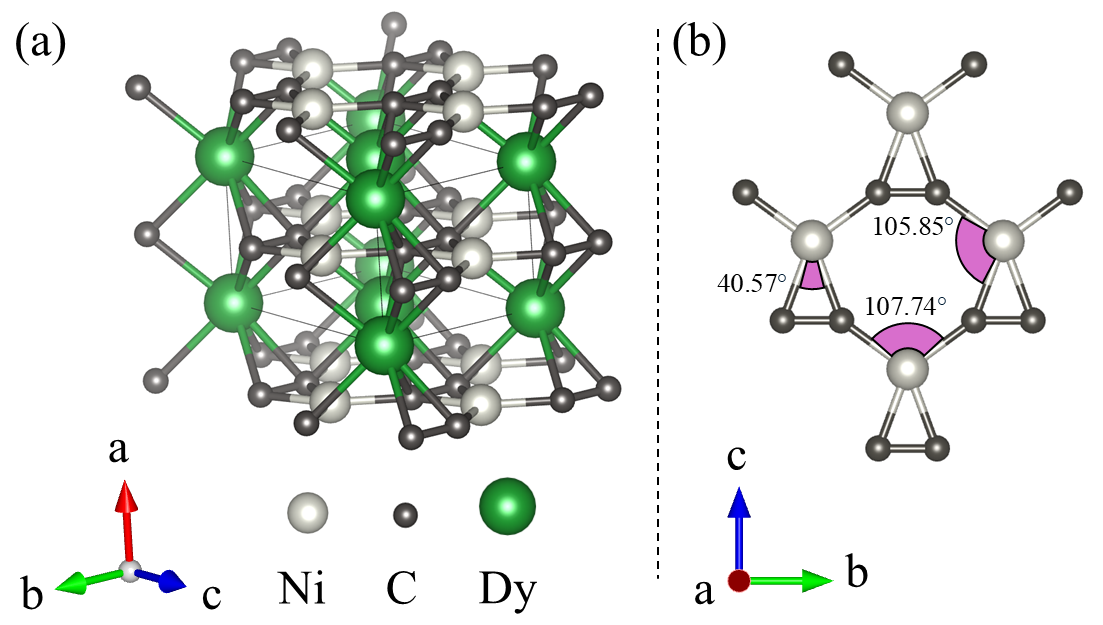}
    \caption{(a) DyNiC$_2$ crystal structure. (b) Face down view on the b-c plane of distorted square planar Ni-C chains.}
    \label{fig:DyNiC2_crystal_structure}
\end{figure}

The DyNiC$_2$ structure is shown in Fig. \ref{fig:DyNiC2_crystal_structure} and has a structure of alternating Ni-C and Dy sheets in the $bc$-plane \cite{jeitschko-1986}. The local Ni-C cluster is a strongly distorted square planar structure with a point group C$_{2v}$. The Ni(II) ion has $n_d = 8$. Low-symmetry clusters, like DyNiC$_2$, experience spin-quenching when the crystal field split levels all lose their degeneracy \cite{pacchioni-1987}, hence the $n_d = 8$ configuration here is fully spin-paired, leading to a spin-0 system. 

Both systems have orthorhombic unit cells and therefore the crystal axes are orthogonal. For describing the direction of propagation and polarization of the x-ray emission we will use the $x$, $y$, $z$ coordinate convention with the obvious mapping to the $a$, $b$, $c$ crystal axes.

% LiVCuO$_4$ was acquired from the Institute of Geology and Mineralogy of the University of Cologne. The crystal was grown from  LiVO$_3$ flux cooled by 0.1K/hr from $\sim$880K as described in Grams \textit{et al.} \cite{grams-2022}. Single crystal X-ray diffraction was performed to confirm the structure and to orient the crystal. 

% DyNiC$_2$ was acquired from the Institute of Solid State Physics TU Wien, Wiedner Hauptstrasse, Wien Austria. The crystal was grown and characterized following the procedure described in Roman \textit{et al.} \cite{roman-2023}. To summarize, the crystal was synthesized using pure elements and then a single crystal was grown using the floating zone technique. It was oriented with Laue method and then characterized by scanning electron microscopy, powder x-ray diffraction to confirm crystal growth and homogeneity.

\section{Methods}
\label{sec:methods}

\subsection{Experimental Setup}
\label{sec:expr_setup}

\begin{figure}
    \centering
    \includegraphics[width=1\linewidth]{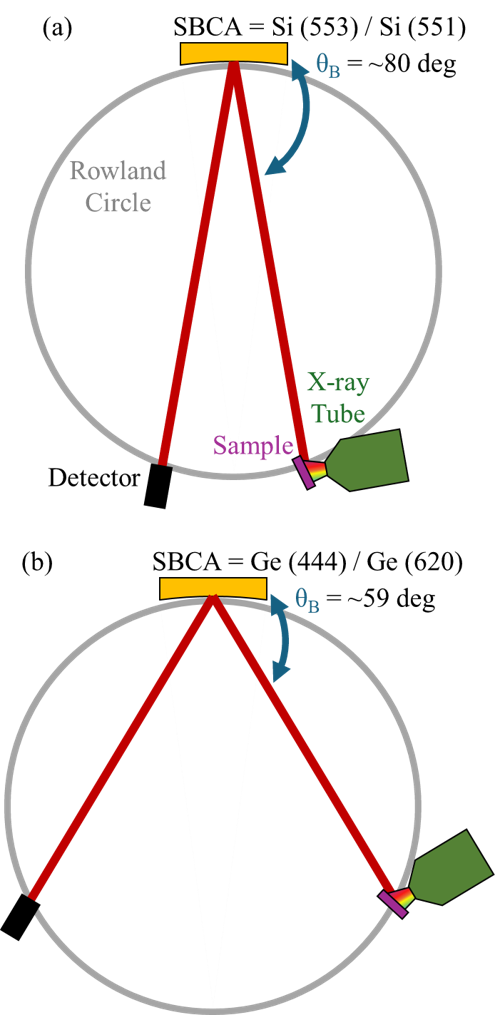}
    \caption{Rowland Circle geometry for unpolarized, (a), and partially polarized, (b), CtC and VtC K$\beta$ XES measurements. For each geometry the first SBCA is for Cu measurements and the second is for Ni measurements.}
    \label{fig:RowlandGeom}
\end{figure}

All XES measurements were performed at room temperature on a laboratory spectrometer described in Jahrman et al. \cite{jahrman-2019}, using a 100W x-ray source with a Pd anode, 10-cm diameter spherically bent crystal analyzer (SBCA), and an Amptek X-123 silicon drift detector on a 1-m Rowland circle. The x-ray tube was held at 2.8 mA current and 35 kV accelerating potential. A 1-mm entrance slit and a mask that covers the outer 30 mm on either side of the SBCA were used to reduce the experimental broadening. Measurements were made with 0.25 eV steps around the features of interest and 1 eV steps outside this region for background determination. Each spectrum was collected over multiple scans, with the specific number of scans chosen to obtain a total of at least 10,000 counts for the K$\beta_{1,3}$ peak and $\sim$1000 counts for the K$\beta_{2,5}$ peak. The first and last few scans were compared and showed no evidence of beam damage.

% \begin{table}[ht]
% \begin{center}
% \begin{tabular}{|c|c|c|c|c|} 
%  \hline
%  \multicolumn{2}{|c|}{Emission} & \multicolumn{3}{|c|}{Spectrometer} \\ \hline
%  Emission Line & Polarized & SBCA & $\theta_B$ (deg.) & Pol. Factor\\ \hline
%  Cu K$\beta$ & No & Si (553) & 80 & 0.88\\ \hline
%   Cu K$\beta$ & Partial & Ge (444) & 58 & 0.19\\ \hline
%  Cu VtC & No & Si (553) & 78 & 0.84\\ \hline
%   Cu VtC & Partial & Ge (444) & 58 & 0.19\\ \hline
%  Ni K$\beta$ & No & Si (551) & 81 & 0.91\\ \hline
%   Ni K$\beta$ & Partial & Ge (620) & 57 & 0.17\\ \hline
%  Ni VtC & No & Si (551) & 80 & 0.88\\ \hline
%  Ni VtC & Partial & Ge (620) & 57 & 0.17\\ \hline
% \end{tabular}
% \caption{Emission line, polarization classification, analyzer, Bragg angle, and polarization factor (R$_p$/R$_s$) for all data presented in this work. See Fig.\ref{fig:RowlandGeom} for graphical representations.}
% \label{tab:spectrometer_setup}
% \end{center} 
% \end{table}

\begin{table}[ht]
\begin{center}
\begin{tabular}{|c|c|c|c|c|} 
 \hline
 \multicolumn{2}{|c|}{Emission} & \multicolumn{3}{|c|}{Spectrometer} \\ \hline
 \shortstack{Emission\\Line} & Polarized & SBCA & $\theta_B$ (deg.) & \shortstack{Pol.\\[0.7ex]Factor} \\ \hline
 Cu K$\beta$ & No & Si (553) & 80 & 0.88\\ \hline
 Cu K$\beta$ & Partial & Ge (444) & 58 & 0.19\\ \hline
 Cu VtC & No & Si (553) & 78 & 0.84\\ \hline
 Cu VtC & Partial & Ge (444) & 58 & 0.19\\ \hline
 Ni K$\beta$ & No & Si (551) & 81 & 0.91\\ \hline
 Ni K$\beta$ & Partial & Ge (620) & 57 & 0.17\\ \hline
 Ni VtC & No & Si (551) & 80 & 0.88\\ \hline
 Ni VtC & Partial & Ge (620) & 57 & 0.17\\ \hline
\end{tabular}
\caption{Emission line, polarization classification, analyzer, Bragg angle, and polarization factor (R$_p$/R$_s$) for all studies presented in this work. See Fig.~\ref{fig:RowlandGeom} for graphical representations.}
\label{tab:spectrometer_setup}
\end{center} 
\end{table}

% , slightly less than unity, and air, slightly more than unity, along with the angle of incidence, the Bragg angle. 

\begin{figure}
    \centering
    \includegraphics[width=1\linewidth]{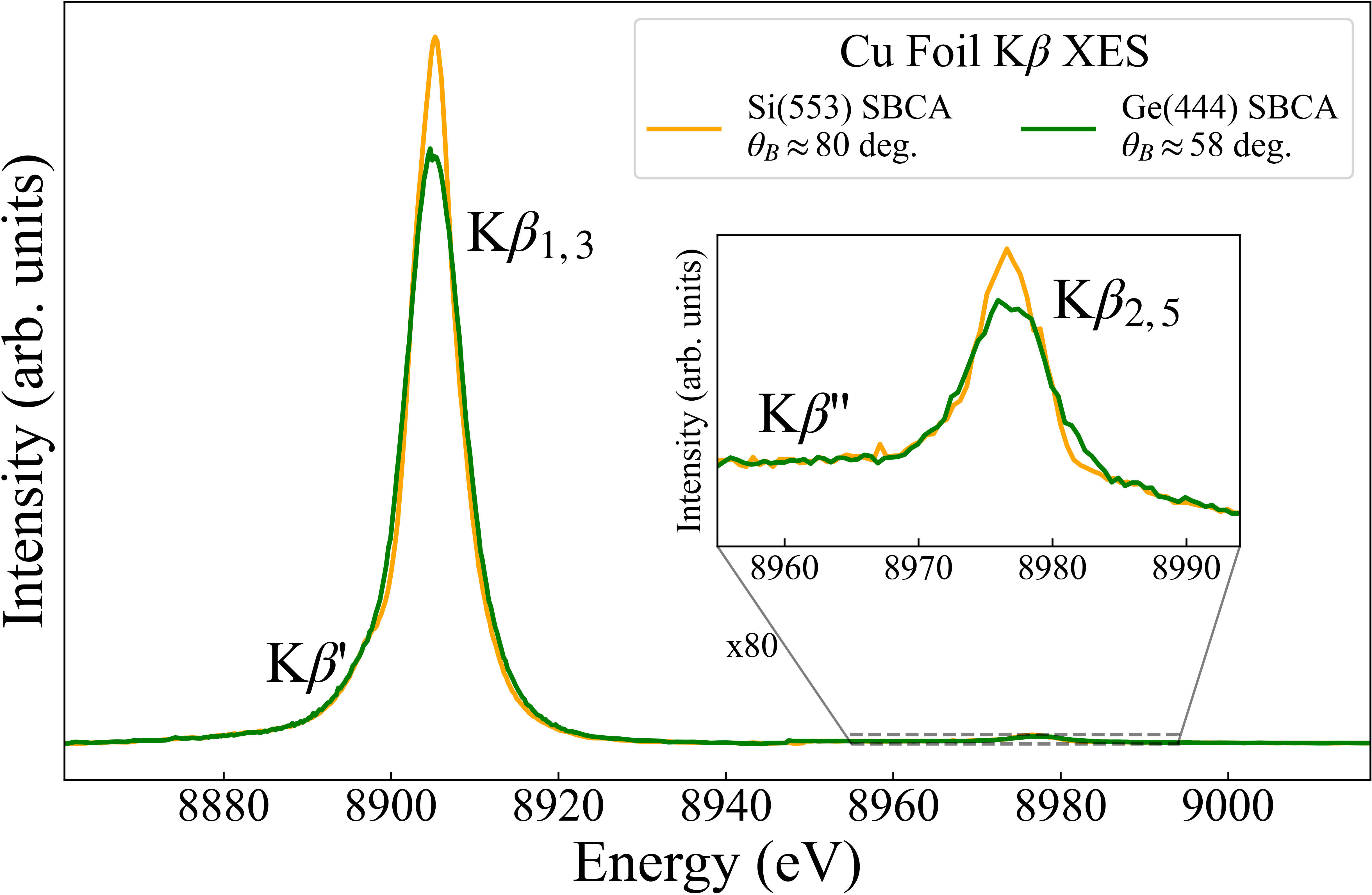}
    \caption{Cu foil K$\beta$ x-ray emission spectra taken at high (orange) and low (green) Bragg angle.}
    \label{fig:high_vs_low_bragg}
\end{figure}

As shown in Fig. \ref{fig:RowlandGeom}, two spectrometer geometries were used to obtain the unpolarized and partially polarized XES measurements, using high and low Bragg angles, respectively. Specific Bragg angles are shown in Table \ref{tab:spectrometer_setup}. All SBCA were used in symmetric reflection geometries. The experimental broadening is $\sim$3-eV greater for the low Bragg angle geometry compared to the high Bragg angle, shown for reference spectra in Fig. \ref{fig:high_vs_low_bragg}, due to source size and Johann error broadening being worse for smaller Bragg angles \cite{chen-2025,bergmann-cramer-1998,rovezzi-2017}. 

\begin{figure}
    \centering
    \includegraphics[width=1\linewidth]{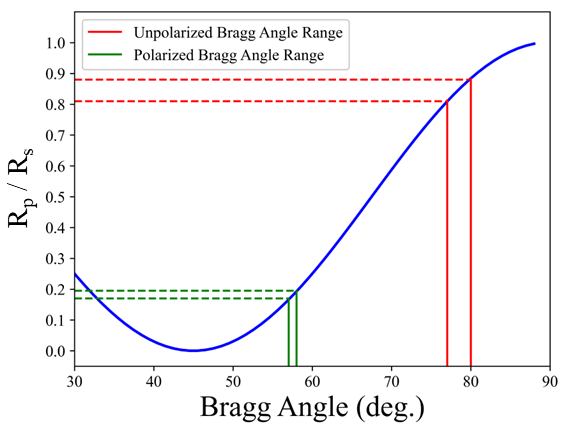}
    \caption{The calculated ratio of reflectance of x-rays with polarization parallel to the Rowland plane (R$_p$) verse perpendicular to the Rowland plane (R$_s$) as a function of Bragg angle for the SBCA. The minimum and maximum Bragg angles for the unpolarized (red) and partially polarized (green) XES spectra are marked.}
    \label{fig:Rp_to_Rs_ratio}
\end{figure}

Polarization control was achieved by exploiting the difference in reflectivity of emission from photons polarized perpendicular (s-polarized) versus parallel (p-polarized) to the reflection plane, see Fig. \ref{fig:Rp_to_Rs_ratio}. The reflectivity of these two polarizations is calculated with Fresnel's equations \cite{hecht2016optics} using the index of refraction for Si or Ge at x-ray energies. This gives a ratio of emission reflecting off the SBCA with polarization parallel (R$_p$) versus perpendicular (R$_s$) to the Rowland plane of $\sim$0.2 for low Bragg angles and $\sim$0.85 for high Bragg angles. Therefore using the spectrometer at low Bragg angles achieves selection of x-rays that are primarily polarized perpendicular to the Rowland plane, meaning that we detect a partially polarized spectra. By orienting the emitting sample with a chosen crystallographic axis perpendicular to the Rowland plane the partially polarized spectra show the transition intensity along the specific axis.

\subsection{Samples and Sample Orientation}
\label{sec:samples}

% Photographs of the two samples are shown in Fig. \ref{fig:LiVCuO4_orientation} (a) and (b). Both samples are plate-like having 4-mm to 8-mm spatial extent in their planar directions and are thinner ($\sim$1-mm) in the perpendicular direction. The LiVCuO$_4$ crystal was grown from  LiVO$_3$ flux cooled by 0.1K/hr from $\sim$880K as described in Grams \textit{et al.} \cite{grams-2022}. Single crystal X-ray diffraction was performed to confirm the structure and to orient the crystal. The DyNiC$_2$ crystal was grown and characterized following the procedure described in Roman \textit{et al.} \cite{roman-2023}. The crystal was synthesized using pure elements and then a single crystal was grown using the floating zone technique. It was oriented with Laue method and then characterized by scanning electron microscopy, powder x-ray diffraction to confirm crystal growth and homogeneity.

\begin{figure}
    \centering
    \includegraphics[width=1\linewidth]{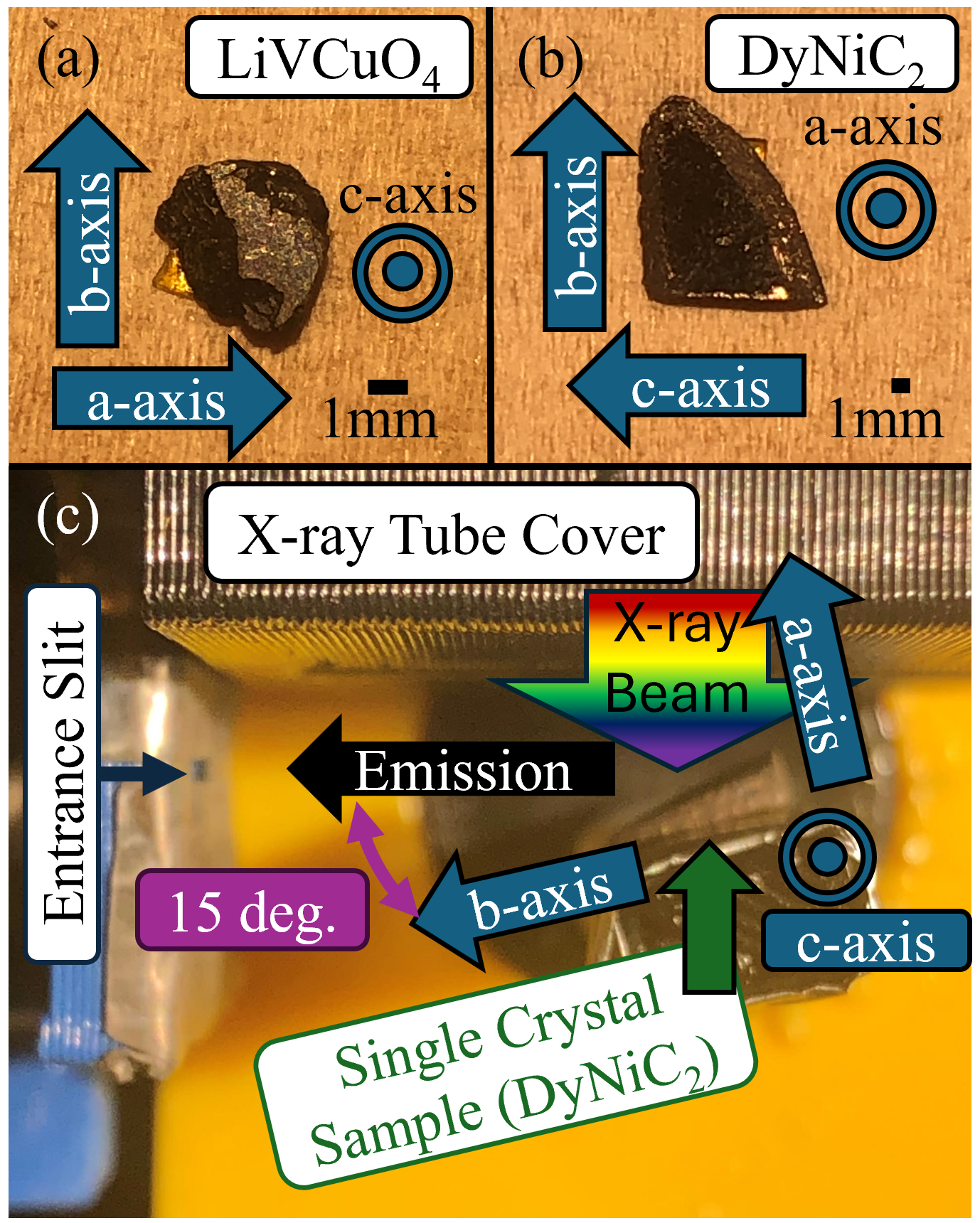}
    \caption{(a) LiVCuO$_4$ single crystal sample. (b) DyNiC$_2$ single crystal sample. (c) DyNiC$_2$ sample shown positioned in the spectrometer sample environment for $z$-polarized, $y$-directed ($\sigma_z$, I$_y$) XES spectra.}
    \label{fig:LiVCuO4_orientation}
\end{figure}

Photographs of the two samples are shown in Fig. \ref{fig:LiVCuO4_orientation} (a) and (b). Both samples are plate-like having 4-mm to 8-mm spatial extent in their planar directions and are thinner ($\sim$1-mm) in the perpendicular direction. The LiVCuO$_4$ crystal was grown from  LiVO$_3$ flux cooled by 0.1K/hr from $\sim$880K as described in Grams \textit{et al.} \cite{grams-2019}. Single crystal X-ray diffraction was performed to confirm the structure and to orient the crystal. The DyNiC$_2$ crystal was grown and characterized following the procedure described in Roman \textit{et al.} \cite{roman-2023}. The crystal was synthesized using pure elements and the floating zone technique. It was oriented with Laue method and then characterized by scanning electron microscopy, powder x-ray diffraction to confirm crystal growth and homogeneity.

The samples were oriented in the spectrometer on a custom 3D printed mount for each measurement so one crystallographic axis is perpendicular to the Rowland plane and another is 15 degrees from the emission direction. An example of this mounting is given in Fig.\ref{fig:LiVCuO4_orientation} (c). The 15 degree offset allows for a much higher count rate with a small correctable loss of directionality along the in-plane crystal axis, discussed in Section \ref{sec:dipole_spectra_extraction}.

% Reference Cu and Ni foils \textbf{FIND AND WRITE WHERE THEY WERE FROM} were used during the calibration process, Section \ref{sec:data_process}

Reference Cu and Ni foils that were used during the calibration process were procured from ESPI metals being 99.995\% elementally pure. 

\subsection{VtC-XES Computational Details}
\label{sec:vtc_calc}

To calculate the VtC-XES and the \textit{l},\textit{m} projected density of states (DOS) we use the real-space Green's function code FEFF10 \cite{kas-2021} which calculates a single-particle Green's function where many-body interactions are approximated via the LDA exchange-correlation potential. This is sufficient for valence level spectroscopies due to the more delocalized nature of the orbitals involved in bonding, and has been demonstrated to perform similarly to time-dependent DFT approaches for VtC-XES \cite{jahrman-2020}. The $x$, $y$, and $z$ polarizations are calculated using the POLARIZATION card and we include both electric dipole and quadrupole transitions with the MULTIPOLE card. While the main contribution to the VtC-XES of 3\textit{d} transition metals comes from the \textit{p} DOS from the dipole transition, roughly 10\% of the intensity can come from quadrupole transitions from the \textit{s} and \textit{d} DOS, as demonstrated in Mortensen \textit{et al.} \cite{mortensen-2017} and in Supplemental Information section \ref{app:moreFEFF}.

The potentials and densities were calculated with the self-consistent field (SCF) approach and an SCF radius of 5.0 \AA\ around the emitting atom and the spectra were calculated with a full multiple scattering (FMS) radius of 7.0 \AA. The XES was calculated using the standard practice of omitting the core hole in accordance with the final state rule. The lifetime broadening from the core hole is already included within FEFF spectra, and a 1.0 eV FWHM Gaussian broadening was convolved with the spectra for comparison with experiment. Spectra were energy shifted independently to align with experiment, which is necessary to account for limitations from the muffin-tin potentials used within FEFF \cite{mortensen-2017}. An extended version of the FEFF code was use to calculate the real spherical harmonic projected density of states, which can be directly compared to the $x$, $y$, $z$ polarized spectra.

\subsection{Data Processing Procedure}
\label{sec:data_process}

\subsubsection{CtC-XES}

\begin{figure}
    \centering
    \includegraphics[width=1\linewidth]{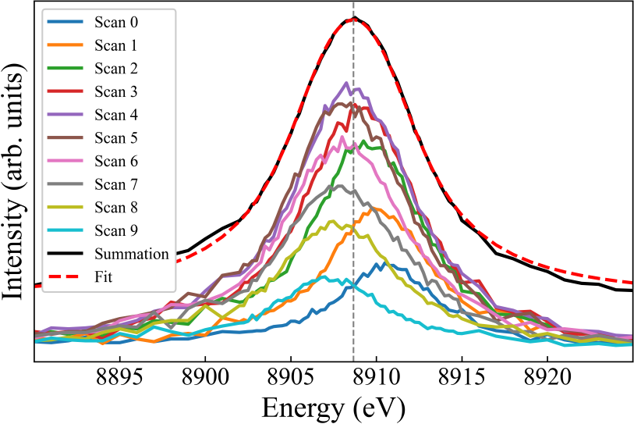}
    \caption{K$\beta$ XES scans of the LiVCuO$_4$ single crystal sample as it is stepped across the entrance slit and the summation of these scans. A vertical gray dashed line marks the center position of the K$\beta_{1,3}$ peak, found by a fit to the summation. This energy is used to apply a Bragg angle correction to the sample spectra, positioning it correctly in energy relative to a Cu foil reference spectra.}
    \label{fig:sample_alignment}
\end{figure}

The CtC K$\beta$ spectra were processed by first averaging all spectra for one sample orientation. A background subtraction calculated from the first and last 5 eV of the spectrum was applied, followed by an integral normalization over the background subtracted spectrum. A unique Bragg angle correction is applied to the spectrum to position it in energy relative to a foil reference spectra. This is necessary because of the sensitivity of the energy scale to the sample position's relative to the entrance slit when the sample is not large enough to fully illuminate the entrance slit, a limitation that is related to the issue addressed in Abramson et al. \cite{abramson-2021}. Following that work, we find that summing a series of alignment scans as the sample steps across the entrance slit creates an effectively-large sample spectrum that fully illuminates the entrance slit. This set of scans can be used to finely calibrate the energy scale for the final measurement with a stationary, small sample - see Fig. \ref{fig:sample_alignment}. The calibration procedure involves comparing the resulting effectively-large sample spectrum's and the normalized, background subtracted spectrum's K$\beta_{1,3}$ energy to find the Bragg angle offset that corrects for the actual illumination of the entrance slit. This step is followed by the usual Bragg angle correction applied equally to all reference and sample spectra to align the Cu and Ni foil reference K$\beta_{1,3}$ energy with their published, standard values. This process aligns both the partially polarized and unpolarized spectra on the same energy scale for comparison.  

\subsubsection{VtC-XES}
\label{sec:data_process_VtC}

VtC-XES spectra were processed by summing all scans for one orientation then subtracting a constant background calculated by averaging the highest 5 eV of data. This was followed by normalization of the VtC region with the integral of the K$\beta_{1,3}$ and K$\beta'$ features to bring the VtC spectra to a consistent molar scale \cite{dhakal-2023}. The Bragg angle corrections determined by the CtC K$\beta$ XES are similarly applied to the VtC-XES. 
%Finally, ray tracing calculations of the two spectrometer geometries, high and low Bragg angle, found a ratio of 1.5 for K$\beta_{2,5}$ detection efficiency when K$\beta_{1,3}$ detection efficiency is set to 1. To compare VtC measurements a linear, Bragg angle dependent intensity correction is therefore applied.

%Finally, a Bragg angle-dependent intensity correction is performed to allow comparison between unpolarized and partially polarized measurements. Sagittal defocusing, and therefore spectrometer detection efficiency, is dependent on Bragg angle, being focused at high Bragg angles and spread at low Bragg angles. Additionally, to cover the same energy range the partially polarized measurements have a small spread in Bragg angle and the unpolarized measurements have a large spread in Bragg angle. These combine to lead to a ratio of detection efficiency between K$\beta_{1,3}$ and K$\beta_{2,5}$ of 1.4 for unpolarized and 0.9 for partially polarized spectra, calculated using ray tracing and spectrometer geometry. These ratios are turned into a linear, Bragg angle dependent intensity correction applied to each spectra.  

\subsubsection{Polarized Spectra Extraction}
\label{sec:dipole_spectra_extraction}

X-ray emission propagating in a particular direction (ex: $z$) is a combination of emission from predominantly dipole transitions in the plane perpendicular to this direction of propagation ($x$ and $y$). For the Rowland circle spectrometer, the direction of propagation is towards the SBCA, and the two polarizations which can reach the detector are orthogonal to the propagation direction: in the Rowland plane ($p$) and perpendicular to the Rowland plane ($s$). The ratio of $s$ and $p$ polarizations which reach the detector depend on the Bragg angle, laid out in Fig. \ref{fig:Rp_to_Rs_ratio} and Table \ref{tab:spectrometer_setup}. Due to the 15 degree difference between crystallographic axis and emission direction in the Rowland plane (Section \ref{sec:crystalsystems}) the $p$-polarization has contribution from polarized spectra along two crystallographic axis while $s$-polarization is only from the polarized spectra of the out of plane axis, example shown in \ref{fig:LiVCuO4_orientation}(c). 

By measuring the emission spectra from a sample along three linearly independent directions, a system of equations is created where each directed spectra is an linear combination of the underlying polarized spectra. The relationship between measured intensities and the underlying polarized emission along the x-, y-, and z- directions is given in Eq. \ref{eq:extraction}. The linear combination coefficients, $A$, have a geometric component due to the 15 degree difference between crystallographic axis and emission direction in the Rowland plane (Section \ref{sec:crystalsystems}) and a reflectivity component due to the R$_p$/R$_s$ ratio of the Rowland circle geometry. $A$ in Eq. \ref{eq:coeff} is given for DyNiC$_2$ with the second row matching the sample orientation shown in Fig. \ref{fig:LiVCuO4_orientation}(c). With $A$ being known, we extract extract the polarized spectra by matrix inversion.

At high Bragg angle, R$_p$/R$_s$ is $\sim$0.85, requiring the extraction procedure of Eq. \ref{eq:extraction} to infer the polarized spectra from the measured unpolarized directional spectra. At low Bragg angle R$_p$/R$_s$ is $\sim$0.2 which heavily favors the out of plane polarization. Performing the dipole extraction on the measured partially polarized directional spectra changes the integral intensity by $\sim$5 percent, meaning that the measured partially polarized directional spectra are approximately equal to the polarized spectra. An important note here here is that this extraction procedure will only be exact for dipole transition spectra. Because the spectra we study in this work are all highly dipole dominated (see Figures \ref{fig:LiVCuO4_VtC-XES_Theory} and \ref{fig:DyNiC2_VtC-XES_Theory}), we will only consider the previously mentioned extraction procedure.
%, meaning that the measured partially polarized directional spectra are approximately equal to the polarized spectra without any additional processing. An important note here here is that this extraction procedure will only be exact for dipole transition spectra. Because the spectra we study in this work are all highly dipole dominated (see Appendix \ref{app:moreFEFF}), we will only consider the previously mentioned extraction procedure.

\begin{subequations}
\label{eq:extraction}
\begin{equation}
    \label{eq:lincomb}
    \begin{bmatrix}
        I_{x}\\I_{y}\\I_{z}
    \end{bmatrix}
    =A
    \begin{bmatrix}
        \sigma_{x}\\\sigma_{y}\\\sigma_{z}
    \end{bmatrix}
\end{equation}
\begin{equation}
    \label{eq:coeff}
    A=
    \begin{bmatrix}
        R_{p}\sin{15\degree}&R{_s}&R{_p}\cos{15\degree}\\
        R{_p}\cos{15\degree}&R{_p}\sin{15\degree}&R{_s}\\
        R{_s}&R{_p}\cos{15\degree}&R{_p}\cos{15\degree}
    \end{bmatrix}
\end{equation}
\end{subequations}

\section{Results and Discussion}
\label{sec:res_and_disc}

\subsection{CtC-XES}
\label{sec:res_and_disc_CtC}

\begin{figure}
    \centering
    \includegraphics[width=1.0\linewidth]{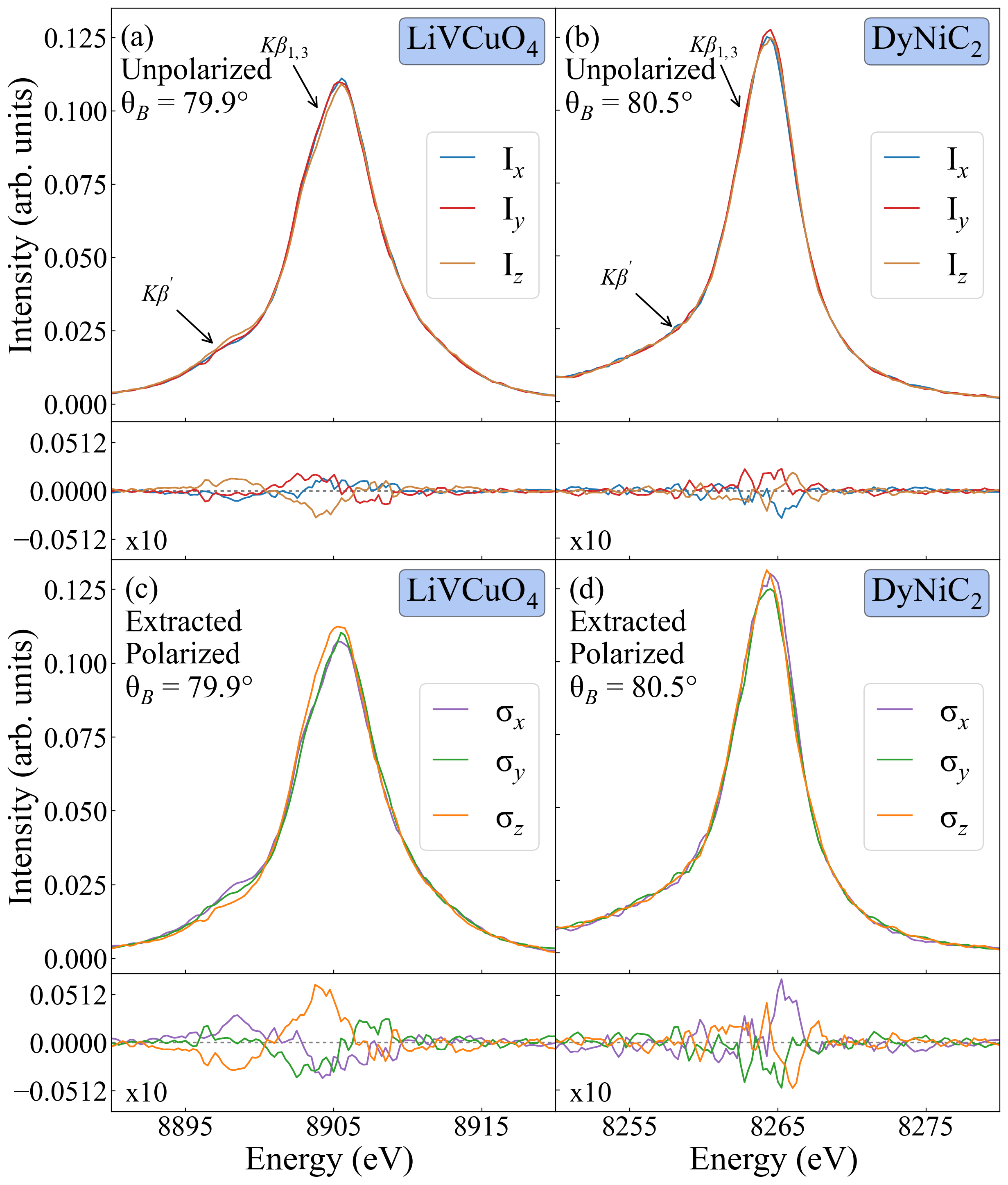}
    \caption{(left) LiVCuO$_4$ Cu and (right) DyNiC$_2$ Ni K$\beta$ CtC-XES. (a, b) Unpolarized emission at high Bragg angle from radiation propagating along the $x$ (blue), $y$ (red), and $z$ (brown) directions. (c, d) polarized spectra extracted from the directional spectra in (a, b) with polarizations along the $x$ (purple), $y$ (green), and $z$ (orange) directions. Difference curves are shown at the bottom of each subplot and are calculated by subtracting the average spectra from each curve. Note the polarization dependence in the K$\beta$' feature for LiVCuO$_4$.}
    \label{fig:Kb-XES_expr}
\end{figure}

We expect core-to-core transitions to demonstrate relatively weak polarization sensitivity given that local environmental effects are only reflected through a coupling between the core hole and valence level, as we discussed in Section \ref{sec:toymodel}. The size of the effect is demonstrated through experimental spectra in Fig. \ref{fig:Kb-XES_expr}. The left and right columns show the K$\beta$ emission from LiVCuO$_4$ and DyNiC$_2$ respectively. The top (a, b) row shows the unpolarized spectra and the bottom (c, d) row shows the extracted polarized spectra along the crystallographic directions using the procedure laid out in Section \ref{sec:dipole_spectra_extraction}.

While the overall anisotropic signals in both CtC K$\beta$ spectra are weak, one key observation is the polarization dependence of the K$\beta$' feature. The spin-1/2 Cu in LiVCuO$_4$ shows a significant difference between the $\sigma_z$ spectrum and the $\sigma_x$/$\sigma_y$ spectra. This, along with knowledge of the crystal field symmetry from the C$_{2v}$ Cu cluster, allows us to determine that the unpaired electron and thus the single 3$d$ hole lies in the $d_{xy}$ orbital (using the crystal axes as a basis). 

When the polarization vector is in the $xy$-plane, the interaction between the newly unpaired 3$p$ electron and the unpaired 3$d_{xy}$ electron produces an energy difference between the spin-aligned (triplet) and spin-opposed (singlet) configurations, leading to a more prominent K$\beta'$ satellite. Conversely, when the polarization vector of the emission is along the $z$-axis, the symmetry of the core hole and the unpaired valence electron do not match and the interaction is weaker, leading to a less prominent K$\beta'$ feature and a larger K$\beta$ main peak for the $z$-polarized spectra in Fig. \ref{fig:Kb-XES_expr} (c). These energy shifts of the K$\beta'$ peak are qualitatively consistent with the behavior observed in the toy model and Figure \ref{fig:Kbeta_toy}, reinforcing our underlying physical interpretation.

The Ni in the DyNiC$_2$ system is spin-0 and therefore produces no K$\beta'$ feature. The absence of a singlet-triplet splitting interaction limits the anisotropic signal to small changes in the energy and intensity of the K$\beta_{1,3}$ main peak. This is similar to what we observe in Fig. \ref{fig:Twoelec_toy_model} of the toy model, where having a spin-0 valence suppresses anisotropy in the polarized spectra. The overall weak polarization dependence of the main K$\beta_{1,3}$ is not unexpected given it is a second-order property transferred to the core level through the core-valence Coulomb interaction.

\begin{figure*}
    \centering
    \includegraphics[width=0.8\linewidth]{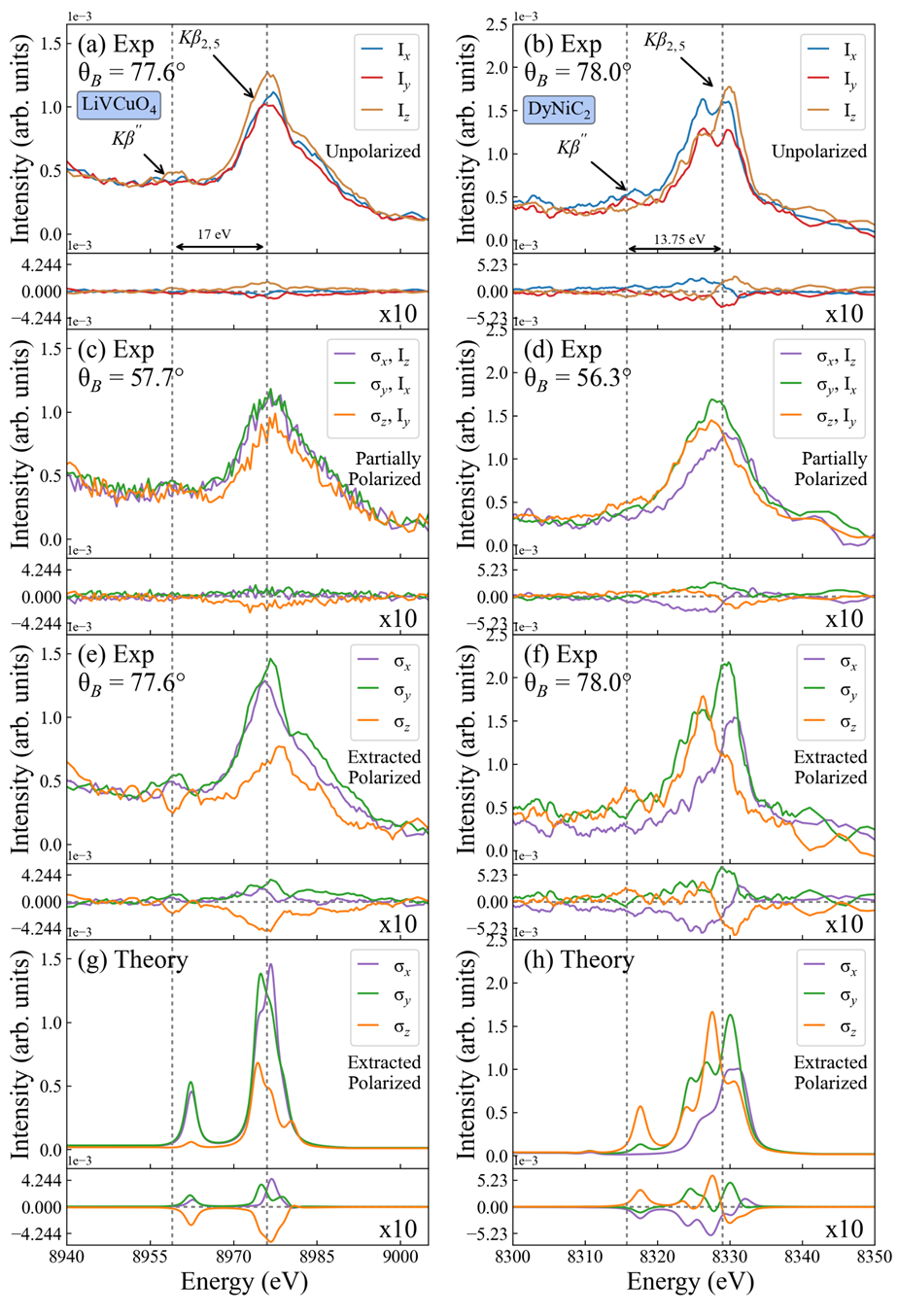}
    \caption{(left) LiVCuO$_4$ Cu and (right) DyNiC$_2$ Ni VtC-XES. Difference curves are shown below each subplot where the difference is taken relative to the averaged (isotropic) spectrum. (a, b) Measured unpolarized emission from radiation propagating along the $x$, $y$, and $z$ directions as defined by the coordinate systems in Section \ref{sec:crystalsystems}. (c, d) Measured partially polarized emission, where each spectrum is dominated by a single polarization axis, albeit with poorer energy resolution. (e, f) Extracted polarized spectra. (g, h) Calculated polarized emission, including both electric dipole and quadrupole components.}
    \label{fig:joined_VtC_theoryvsexpr}
\end{figure*}

\subsection{VtC-XES}

The Cu and Ni VtC emission from LiVCuO$_4$ and DyNiC$_2$ are shown in Fig. \ref{fig:joined_VtC_theoryvsexpr}. Unpolarized (high Bragg angle) and partially polarized (low Bragg angle) are shown in subplots (a, b) and (c, d) respectively. The extracted polarization spectra in subplots (e, f) are calculated from combinations of the directional spectra, as described in the system of equations given in Section \ref{sec:dipole_spectra_extraction}. The theory spectra in subplots (g, h) are calculated according to Section \ref{sec:vtc_calc}, and follow the exact same extraction procedure for the purpose of comparison. 

The directed emission spectra (a, b) show weakly anisotropic behavior, which is expected given that they are averages of two polarization contributions. The partially polarized spectra in (c, d) qualitatively match with the polarized spectra (e, f) and the calculated polarized spectra (g, h). They show clear polarization differences, such as the ability to distinguish which polarized spectra are most contributing to the K$\beta''$ ligand peak or K$\beta_{2,5}$ feature, indicating the usefulness of these spectropolarimeter measurements. But the $\sim$20 percent contribution of in-plane polarized spectra, Fig. \ref{fig:Rp_to_Rs_ratio}, and the lower resolution, Fig. \ref{fig:high_vs_low_bragg}, for the partially polarized measurements restricts its quantitative analysis.
%The directed emission spectra (a, b) show weakly anisotropic behavior, which is expected given that they are averages of two polarization contributions. The extracted polarization spectra in (e, f) qualitatively match the measured partially polarized spectra in (c, d), but with higher energy resolution (Fig. \ref{fig:high_vs_low_bragg}). This provides clearer polarization differences, such as the ability to distinguish which polarized spectra are most contributing to the K$\beta''$ ligand peak.

For LiVCuO$_4$ the nearly square planar CuO$_4$ structure means the Cu-O bonds lie in the $xy$-plane. This geometry means that the Cu $d_{z^2}$ orbital lacks suitable ligand orbitals to form $\sigma$ bonds, leading to reduced electron density along the $z$-axis and, consequently, weaker $z$-polarized ($\sigma_z$) emission compared to the $x$ and $y$ polarizations. The unpolarized $I_z$ emission in (a) is therefore stronger than the $I_x$ and $I_y$ signals, as it averages over stronger in-plane transitions. Similarly, the partially polarized and extracted polarized spectra in (c) and (e) show suppressed $\sigma_z$ contributions, while $\sigma_x$ and $\sigma_y$ remain strong and nearly identical due to the symmetry of the in-plane bonding environment. This also accounts for the presence of the K$\beta''$ ligand peak at 8958 eV in both $\sigma_x$ and $\sigma_y$.

% For DyNiC$_2$ the nonuniform planar structure results in the $y$-polarized dipole emission spectrum being the strongest followed by the $z$ and lastly the $x$ and as a result the directed spectra follow the reverse order for integrated intensity: $I_x$ strongest, then $I_z$, and finally $I_y$. For both systems the $K\beta''$ satellite is relatively weak compared to the main $K\beta_{2,5}$ peak. This can be understood due to the long metal-ligand bond distances, 2.14\AA\ average for Cu-O bonds in the $xy$-plane of LiVCuO$_4$ and 1.96\AA\ average for Ni-C bonds in the $yz$-plane of DyNiC$_2$, resulting in a small overlap of the ligand 2\textit{s} and metal 1\textit{s} orbitals since both orbitals are relatively localized. 

% The measured partially polarized XES spectra in subplots (c, d) are more distinctly anisotropic than the directed with the anisotropy following the expectations from crystal structure. 

For DyNiC$_2$, the distorted planar structure leads to highly anisotropic emission, with the $\sigma_y$ spectrum being the strongest, followed by $\sigma_z$, and then $\sigma_x$, which is the weakest due to the absence of a Ni–C bond along the $x$-direction. This is analogous to the lack of Cu-O bonding along the $z$-axis in LiVCuO$_4$. The distortion of the NiC$_4$ cluster results in non-equivalent $\sigma_y$ and $\sigma_z$ spectra. The differences are consistent with the Ni–C bond angles in the $bc$-plane: an average of 42.78$^\circ$ relative to the $b$/$y$-axis and 37.08$^\circ$ relative to the $c$/$z$-axis. This slight compression along the $z$-axis manifests in the spectra as a $\sim$3 eV energy shift of the $\sigma_z$ K$\beta_{2,5}$ emission peak compared to the $\sigma_y$ one, and a more intense C ligand peak at 8316 eV in the $z$-polarization.

In both DyNiC$_2$ and LiVCuO$4$, the $K\beta''$ satellite is relatively weak compared to the main $K\beta_{2,5}$ peak, a trend attributable to the long metal–ligand bond distances—1.96 $\AA$ for Ni–C bonds in DyNiC$_2$ and 2.14 $\AA$ for Cu–O bonds in LiVCuO$4$ \cite{bergmann1999}. This reduces the hybridization between the ligand 2\textit{s} and metal valence orbitals, thereby suppressing the weak $K\beta''$ feature. Additionally, the $K\beta''$–$K\beta_{2,5}$ energy separation is roughly 5 eV greater in LiVCuO$_4$ than in DyNiC$_2$, reflecting the difference in 2\textit{s} binding energies between O and C ligands \cite{miajaavila-2020}.

The extracted polarization spectra in subplots (e, f) match well with the calculated polarized spectra in subplots (g, h), and their residuals. However, one important difference between the calculated and extracted is that the position of the calculated $K\beta''$ peak is too low in energy by about 2 eV compared to experiment. This is likely due to limitations imposed by the use of muffin-tin potentials within the FEFF code to approximate the scattering potential, which tend to underestimate the anisotropy of the electron density in the interstitial regions. As a result, the hybridization between the transition metal 3$d$ and ligand 2$p$ states may be overestimated, artificially shifting the energy position of the $K\beta''$ peak.

\subsection{Future Directions}

The spectropolarimetry to directly measure the polarized spectra (low Bragg angle) suffered as a analytical measurement from low resolution and only partial polarization sensitivity. Both factors can be greatly improved upon by working in an asymmetric Rowland configuration \cite{gironda-2024}, allowing for a selection of reflection geometry with near 45 degree Bragg angle for perfect polarization sensitivity with emission incident perpendicular to the SBCA face reducing one of two main broadening mechanisms, Johann error \cite{chen-2025}. If the spectropolarimetry measurements were performed with the optimally polarized asymmetric Rowland geometry and with micro-focused synchrotron radiation the other main broadening mechanism, source size error, could be eliminated resulting in a direct, high-fidelity measurement of polarized spectra. 
%In addition to the chemical analysis, we also showed instrumentally that by measuring low Bragg angle polarized emission, individual polarized spectra can be detected with a single analyzer. Due to the spectrometer configuration in this work the low Bragg angle spectra had relatively poor resolution but by working in an asymmetric Rowland configuration polarized emission spectra with high resolution could be directly measured \cite{gironda-2024}. In the absence of an asymmetric Rowland configuration we showed that unpolarized high Bragg angle measurements can be used to extract the underlying polarized measurements while maintaining the resolution the high Bragg angle provides.

An additional benefit of performing the direct polarized measurements at a synchrotron source is the increased flux which allows for resolving weaker spectral features such as quadrupole transitions, which depend on both the direction of propagation and the polarization axis, adding an extra layer of complexity. Because quadrupole transitions are much weaker than the dominant dipole transitions (see Supplemental Information \ref{app:moreFEFF}), their detection requires careful analysis of intensity variations with crystal orientation. Lower experimental broadening and increased flux would make it significantly easier to isolate and identify these subtle features.
%This would be particularly useful for resolving weaker spectral features such as quadrupole transitions, which depend on both the direction of propagation and the polarization axis, adding an extra layer of complexity. Because quadrupole transitions are much weaker than the dominant dipole transitions (see Section \ref{app:moreFEFF}), their detection requires careful analysis of intensity variations with crystal orientation. Achieving lower experimental broadening would make it significantly easier to isolate and identify these subtle features.

\section{Conclusion}
\label{sec:conclusion} 

We have shown that the anisotropy observed in the polarized CtC and VtC-XES emission spectra reflects the local asymmetry around the 3$d$ transition metal in single crystal systems. The polarized emission along a specific crystallographic direction is obtained directly via partially polarized measurements using single crystal spectropolarimetry and indirectly with a new technique to extract polarized spectra from unpolarized emission. The polarized emission provided information about the symmetry of the local electronic structure, allowing us to determine the anisotropic distribution of the occupied magnetic orbitals. By combining these results with a polarized absorption study, such as an XLD analysis of the Cu and Ni K-edges, our findings could be further corroborated by looking for a polarization dependent pre-edge feature. In this way, combining both absorption and emission techniques would provide a more complete understanding of the orbital anisotropy and would help disentangle the full anisotropic electronic structure of complex materials.

%By resolving the emission polarization along specific crystallographic directions, we obtained direct information about the local electronic structure, particularly the anisotropic distribution of unoccupied states.

The CtC K$\beta$ features showed clear polarization dependence for Cu in LiVCuO$_4$, which can be attributed to the spin-1/2 nature of Cu$^{2+}$. This allows for Coulomb coupling between the 3\textit{p} and 3\textit{d} orbitals. The emission weight shifts toward the K$\beta'$ feature when the polarization is in the ligand plane and toward the K$\beta_{1,3}$ feature when it is perpendicular to the plane. In contrast, DyNiC$_2$ contains Ni$^{2+}$ in a spin-0 configuration, and its CtC spectrum showed little to no polarization dependence. This contrasting behavior is consistent with predictions from a simple toy model.

% The CtC K$\beta$ features demonstrated polarization dependence for Cu due to it being spin-1/2, allowing coupling between the 3\textit{p} and 3\textit{d} orbitals. Weight was shifted to the K$\beta'$ feature when the emission polarization was in the ligand plane and shifted to the K$\beta_{1,3}$ feature when the emission polarization was perpendicular to the planar structure. The Ni spectra showed little to no polarization dependence in the CtC K$\beta$ due to it being a spin-0 system. Both materials matched the toy model behavior.

% The VtC K$\beta$ features showed strong polarization dependence in both materials. The K$\beta''$ and the K$\beta_{2,5}$ features were more intense when the polarization was along a ligand direction, with the shape of the K$\beta_{2,5}$ feature also changing more with the local electronic structure. This was matched qualitatively by FEFF calculations.

In the VtC region, both systems exhibited strong polarization effects. The K$\beta''$ and K$\beta_{2,5}$ features became more intense when the polarization aligned with metal-ligand bond directions, consistent with expectations based on orbital occupation. Additionally, the K$\beta''$ peak entirely disappeared for out-of-plane polarizations, further emphasizing the directional nature of the metal-ligand bonding. These trends were qualitatively reproduced by FEFF calculations, reinforcing the interpretation that VtC-XES is sensitive to directional bonding and orbital interactions \cite{jahrman-2020}.

Overall, these results establish polarization-resolved XES as a tool for probing anisotropic electronic environments in transition metal systems. This technique provides access to subtle variations in the character of the occupied orbitals, which could be leveraged in future studies to resolve weak quadrupole transitions and characterize ligand field asymmetry. Just as conventional XES complements conventional XAS by probing the occupied density of states, polarized XES serves as a natural complement to polarized XAS, offering a more complete picture of directionally resolved electronic structure.

\section{Acknowledgments}

JJK and JJR acknowledge support from the Theory Center for Materials and Energy Spectroscopies (TIMES) at SLAC funded by DOE BSE Contract DE-AC02-76SF00515. CAC was supported by the National Science Foundation Graduate Research Fellowship Program under Grant No. DGE-2140004. GTS and JEA were supported by funding from the U.S. Department of Energy in the Nuclear Energy University Program under Contract No. DE-NE0009158.  Any opinions, findings, and conclusions or recommendations expressed in this material are those of the author(s) and do not necessarily reflect the views of the National Science Foundation or the U.S. Department of Energy. PB acknowledges funding by the DFG (German Research Foundation) via Project No. 277146847-CRC 1238 (subproject A02). Financial support for M.R. by Grant No. BPN/BEK/2021/1/00245/DEC/1 of the Bekker Program of the Polish National Agency for Academic Exchange (NAWA) is gratefully acknowledged.

\newpage
\section{Appendix}
\appendix
\section{Directional and Polarization Dependence of Emitted Radiation}
\label{app:POW_SPEC_APP}

The radiation pattern of any dipole or quadrupole transition element between two hydrogenic orbitals can be calculated by directly integrating the matrix elements from Eq \ref{eq:DipoleApprox} (dipole) and Eq \ref{eq:QuadrupoleApprox} (quadrupole) for an arbitrary $\vec{k}$. The propagation vector $\vec{k}$ and polarization vector $\vec{\epsilon}$ are given in Eq \ref{eq:kandepsilon}, where $\delta$ is the angle $\vec{\epsilon}$ makes in the plane perpendicular to $\vec{k}$ which will be integrated out. This follows the same convention established in \cite{laihia-1998}.

\begin{equation}
    \sigma^{\mathrm{quad}} \propto \sum_f \left| \left\langle f \left| (\hat{\boldsymbol{\epsilon}} \cdot \vec{r}) (\vec{k} \cdot \vec{r}) \right| i \right\rangle \right|^2 \delta(E_f - E_i + \hbar \omega)
    \label{eq:QuadrupoleApprox}
\end{equation}

\begin{subequations} \label{eq:kandepsilon}
    \begin{gather}
        \vec{k}(\theta, \phi) = 
        \begin{bmatrix}
            \sin{\theta} \cos{\phi} \\
            \sin{\theta} \sin{\phi} \\
            \cos{\theta}
        \end{bmatrix} \label{eq:kandepsilon_a} \\
        \vec{\epsilon}(\theta, \phi, \delta) = 
        \begin{bmatrix}
            \cos{\theta} \cos{\phi} \cos{\delta} - \sin{\phi} \sin{\delta} \\
            \cos{\theta} \sin{\phi} \cos{\delta} - \cos{\omega} \sin{\delta} \\
            - \sin{\theta} \cos{\delta}
        \end{bmatrix} \label{eq:kandepsilon_b}
    \end{gather}
\end{subequations}

\begin{figure}
    \centering
    \includegraphics[width=1\linewidth]{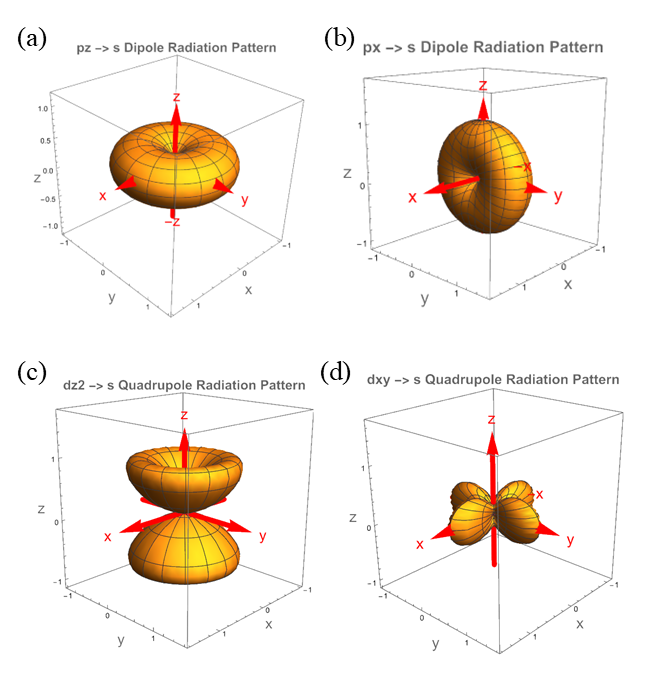}
    \caption{Radiation pattern of dipole and quadrupole transitions between different $l$, $m$ orbitals. (a) and (b) show dipole transitions from the $p_z$ and $p_x$ to $s$ orbitals respectively while (c) and (d) show quadrupole transition from the $d_{z2}$ and $d_{xy}$ to $s$ orbitals. The shape of the radiation patterns are equivalent to oscillating charges with spatial distributions that are consistent with the lobes of positive and negative phases from each atomic orbital.}
    \label{fig:dipandquad}
\end{figure}

For example, the angular component of a dipole transition from a $p_z$ orbital to a $s$ orbital is written out in Eq \ref{eq:pz_to_s}, where $\theta$, $\phi$, and $\delta$ are as defined in equation \ref{eq:kandepsilon}, and $\rho$ and $\omega$ are dummy variables used for evaluating the matrix element. The matrix element ends up simply as $\bra{s} \vec{\epsilon} \cdot \vec{r} \ket{p_z} \propto \sin{\theta}$, and the radiation pattern is given by $(\sin{\theta})^2$, as is shown in Fig. \ref{fig:dipandquad} (a). We also show examples for the $p_x \rightarrow s$, $d_{z2} \rightarrow s$, and $d_{xy} \rightarrow s$. The surface of the radiation pattern gives a qualitative measure of how much radiation is emitted in a particular direction from a given dipole transition element. Given the generalized definitions of $\vec{k}$ and $\vec{\epsilon}$, it's possible to invert them and derive the 'polarization' pattern for a fixed polarization direction and a propagation vector integrated over a plane ($\delta$), which of course produces the exact same patterns but with a different interpretation. 

\begin{align}
    &\bra{s} \vec{\epsilon} \cdot \vec{r} \ket{p_z} = \Bigg[ \int_0^{2\pi} \int_0^{2\pi} \int_0^{\pi} \label{eq:pz_to_s} \\
    & \Big(  (\cos{\theta} \cos{\phi} \cos{\delta} - \sin{\phi} \sin{\delta}) \sin{\rho} \cos{\omega} \Big) \notag \\
    & \Big(  (\cos{\theta} \sin{\phi} \cos{\delta} - \cos{\omega} \sin{\delta}) \sin{\rho} \sin{\omega} \Big) \notag \\
    & \Big(  (-\sin{\theta} \cos{\delta}) \cos{\rho} \Big) (\sin{\rho} \cos{\rho}) \; d\rho \, d\omega \, d\delta \Bigg] \notag \\
    & \quad \quad \quad \quad \quad \propto \sin{\theta} \notag
\end{align}

\section{Additional FEFF calculations}
\label{app:moreFEFF}

\begin{figure}
    \centering
    \includegraphics[width=1\linewidth]{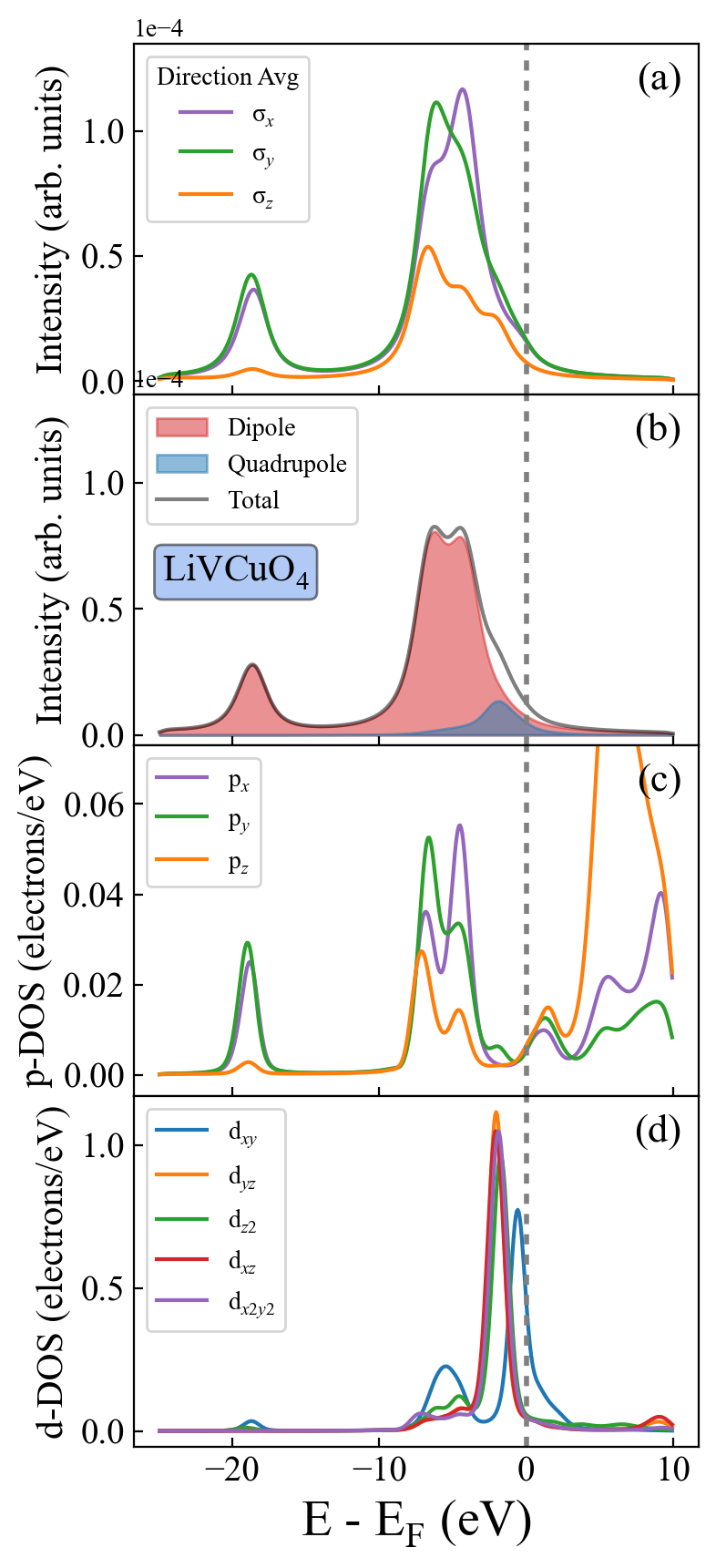}
    \caption{FEFF calculated Cu VtC spectra and DOS from LiVCuO$_4$. (a) Polarized x-ray emission with the quadrupole component averaged over directions perpendicular to the polarization axes. (b) Isotropic emission separated into dipole (red) and quadrupole (blue) components. (c) The p-projected density of states. (d) The d-projected density of states.}
    \label{fig:LiVCuO4_VtC-XES_Theory}
\end{figure}

\begin{figure}
    \centering
    \includegraphics[width=1\linewidth]{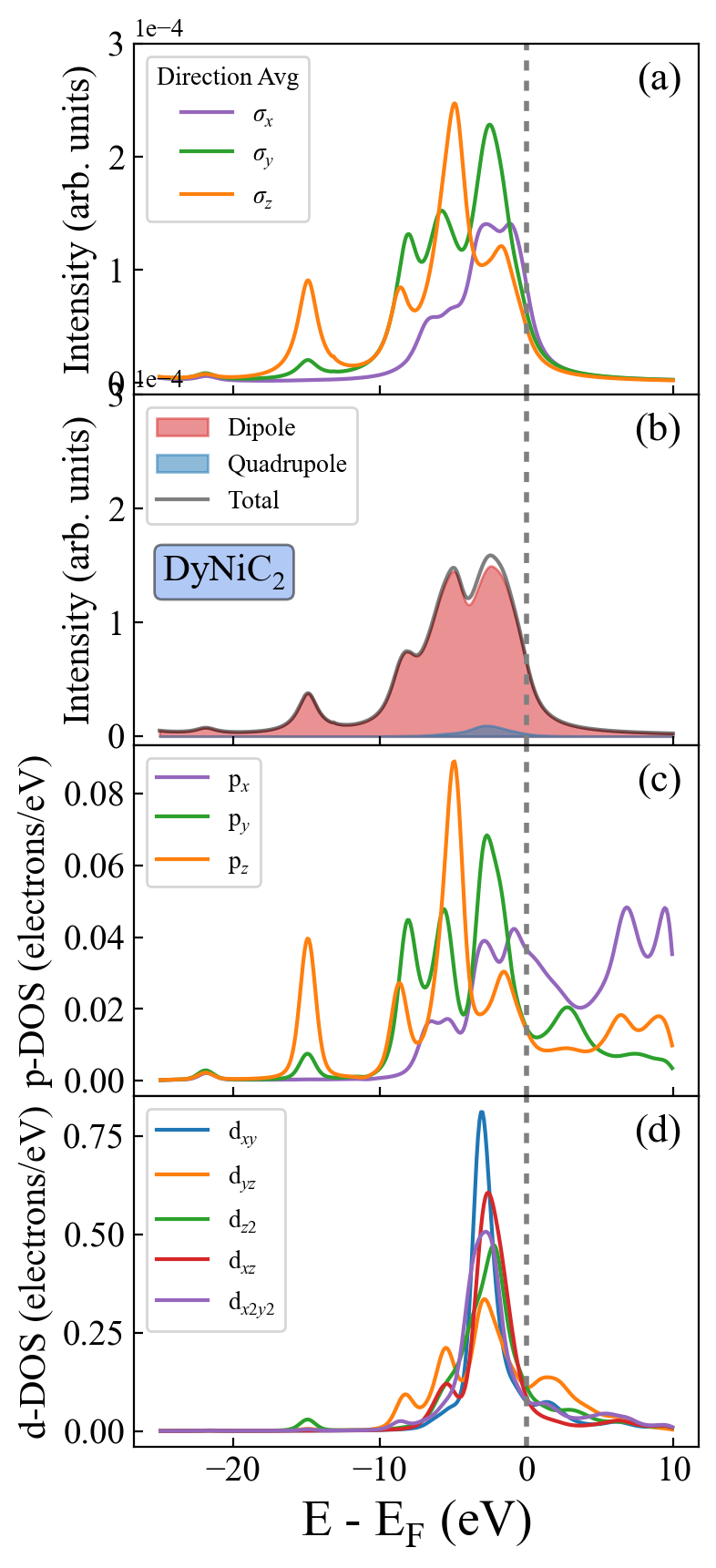}
    \caption{FEFF calculated Ni VtC spectra and DOS from DyNiC$_2$. (a) Polarized x-ray emission with the quadrupole component averaged over directions perpendicular to the polarization axes. (b) Isotropic emission separated into dipole (red) and quadrupole (blue) components. (c) The p-projected density of states. (d) The d-projected density of states.}
    \label{fig:DyNiC2_VtC-XES_Theory}
\end{figure}

X-ray emission is often interpreted as a reflection of the occupied density of states \cite{tetef-2021, lafuerza-2020, glatzel-2004} in the presence of a core hole. We see this represented in figures \ref{fig:LiVCuO4_VtC-XES_Theory} and \ref{fig:DyNiC2_VtC-XES_Theory} which show how the polarized VtC-XES in subplot (a) matches up almost exactly with the p-projected DOS in subplot (c). We also note that the quadrupole contribution shown in subplot (b) is relatively weak, with it making up only 10\%  and 3\% of the total spectral intensity for LiVCuO$_4$ and DyNiC$_2$ respectively. This supports the approximation made when extracting the polarizations from the directional spectra. The main contribution to the quadrupole transition is the $d$-projected DOS which is much larger (in units of electron/eV) than the p-DOS. The overall weak contribution to the total spectra is due to the additional $\vec{k}$ term in the quadrupole matrix element and the small overlap between the metal 3$d$ orbital and 1$s$ orbital.

\newpage
\bibliographystyle{unsrt}
\bibliography{main}

\begin{thebibliography}{10}

\bibitem{evans2021layered}
H.~A. Evans, L.~Mao, R.~Seshadri, and A.~K. Cheetham.
\newblock Layered double perovskites.
\newblock {\em Annual Review of Materials Research}, 51:351--380, 2021.

\bibitem{zhou-2022}
Faran Zhou, Kyle Hwangbo, Qi~Zhang, Chong Wang, Lingnan Shen, Jiawei Zhang, Qianni Jiang, Alfred Zong, Yifan Su, Marc Zajac, Youngjun Ahn, Donald~A. Walko, Richard~D. Schaller, Jiun-Haw Chu, Nuh Gedik, Xiaodong Xu, Di~Xiao, and Haidan Wen.
\newblock {Dynamical criticality of spin-shear coupling in van der Waals antiferromagnets}.
\newblock {\em Nature Communications}, 13(1), 11 2022.

\bibitem{schlappa2012spinorbital}
J.~Schlappa, K.~Wohlfeld, K.~J. Zhou, M.~Mourigal, M.~W. Haverkort, V.~N. Strocov, C.~Monney, T.~Hunt, J.~Stähler, H.~M. Ronnow, et~al.
\newblock Spin–orbital separation in the quasi-one-dimensional mott insulator sr$_2$cuo$_3$.
\newblock {\em Nature}, 485(7396):82--85, 2012.

\bibitem{southworth1991anisotropy}
S.H. Southworth, D.W. Lindle, R.~Mayer, and P.L. Cowan.
\newblock Anisotropy of polarized x-ray emission from molecules.
\newblock {\em Physical Review Letters}, 67(1):1098--1101, 1991.

\bibitem{stohr-1995}
J.~Stöhr.
\newblock {X-ray magnetic circular dichroism spectroscopy of transition metal thin films}.
\newblock {\em Journal of Electron Spectroscopy and Related Phenomena}, 75:253--272, 1995.

\bibitem{van-der-laan-2014}
Gerrit Van Der~Laan and Adriana~I. Figueroa.
\newblock {X-ray magnetic circular dichroism—A versatile tool to study magnetism}.
\newblock {\em Coordination Chemistry Reviews}, 277-278:95--129, 2014.

\bibitem{heppell-2025}
E.~Heppell, F.~Maccherozzi, L.~S.~I. Veiga, S.~Langridge, G.~Van Der~Laan, T.~Hesjedal, and D.~Backes.
\newblock {Handle on the antiferromagnetic spin structure of NiO using a ferromagnetic adlayer}.
\newblock {\em Physical Review Materials}, 9(1):4408, 2025.

\bibitem{sobota-2021}
Jonathan~A. Sobota, Yu~He, and Zhi-Xun Shen.
\newblock {Angle-resolved photoemission studies of quantum materials}.
\newblock {\em Reviews of Modern Physics}, 93(2):5006, 2021.

\bibitem{rosenberg-2022}
Elliott Rosenberg, Jonathan~M. DeStefano, Yucheng Guo, Ji~Seop Oh, Makoto Hashimoto, Donghui Lu, Robert~J. Birgeneau, Yongbin Lee, Liqin Ke, Ming Yi, and Jiun-Haw Chu.
\newblock {Uniaxial ferromagnetism in the kagome metal TbV6Sn6}.
\newblock {\em Physical Review B}, 106(11):5139, 2022.

\bibitem{bergner-2022}
Derek Bergner, Tai Kong, Ping Ai, Daniel Eilbott, Claudia Fatuzzo, Samuel Ciocys, Nicholas Dale, Conrad Stansbury, Drew~W. Latzke, Everardo Molina, Ryan Reno, Robert~J. Cava, Alessandra Lanzara, and Claudia Ojeda-Aristizabal.
\newblock {Polarization dependent photoemission as a probe of the magnetic ground state in the van der Waals ferromagnet VI3}.
\newblock {\em Applied Physics Letters}, 121(18):3104, 2022.

\bibitem{falub-2005}
M.~C. Falub, M.~Shi, P.~R. Willmott, J.~Krempasky, S.~G. Chiuzbaian, K.~Hricovini, and L.~Patthey.
\newblock {Polarization-dependent angle-resolved photoemission spectroscopy study of La1-xSrxMnO3}.
\newblock {\em Physical Review B}, 72(5):4444, 2005.

\bibitem{hajiri-2012}
T~Hajiri, R~Niwa, T~Ito, M~Matsunami, B~H Min, S~Kimura, and Y~S Kwon.
\newblock {Polarization-dependent three-dimensional angle-resolved photoemission study on LiFeAs}.
\newblock {\em Journal of Physics Conference Series}, 391:012125, 2012.

\bibitem{ament-2011}
Luuk J.~P. Ament, Michel Van~Veenendaal, Thomas~P. Devereaux, John~P. Hill, and Jeroen Van Den~Brink.
\newblock {Resonant inelastic x-ray scattering studies of elementary excitations}.
\newblock {\em Reviews of Modern Physics}, 83(2):705--767, 2011.

\bibitem{ulrich-2009}
C.~Ulrich, L.~J.~P. Ament, G.~Ghiringhelli, L.~Braicovich, M.~Moretti Sala, N.~Pezzotta, T.~Schmitt, G.~Khaliullin, J.~Van Den~Brink, H.~Roth, T.~Lorenz, and B.~Keimer.
\newblock {Momentum dependence of orbital excitations in Mott-Insulating titanates}.
\newblock {\em Physical Review Letters}, 103(10):7205, 2009.

\bibitem{harada-2002}
Yoshihisa Harada, Kozo Okada, Ritsuko Eguchi, Akio Kotani, Hidenori Takagi, Tomoyuki Takeuchi, and Shik Shin.
\newblock {Unique identification of Zhang-Rice singlet excitation in Sr2CuO2Cl}.
\newblock {\em Physical Review B, Condensed matter}, 66(16):5104, 2002.

\bibitem{sala-2011}
M~Moretti Sala, V~Bisogni, C~Aruta, G~Balestrino, H~Berger, N~B Brookes, G~M De~Luca, D~Di~Castro, M~Grioni, M~Guarise, P~G Medaglia, F~Miletto Granozio, M~Minola, P~Perna, M~Radovic, M~Salluzzo, T~Schmitt, K~J Zhou, L~Braicovich, and G~Ghiringhelli.
\newblock {Energy and symmetry of dd excitations in undoped layered cuprates measured by Cu L3 resonant inelastic x-ray scattering}.
\newblock {\em New Journal of Physics}, 13(4):043026, 2011.

\bibitem{van-veenendaal-2006}
Michel Van~Veenendaal.
\newblock {Polarization dependence ofL- andM-Edge resonant inelastic X-Ray scattering in Transition-Metal compounds}.
\newblock {\em Physical Review Letters}, 96(11):7404, 2006.

\bibitem{glatzel-2004}
Pieter Glatzel and Uwe Bergmann.
\newblock {High resolution 1s core hole X-ray spectroscopy in 3d transition metal complexes—electronic and structural information}.
\newblock {\em Coordination Chemistry Reviews}, 249(1-2):65--95, 2004.

\bibitem{lafuerza-2020}
Sara Lafuerza, Andrea Carlantuono, Marius Retegan, and Pieter Glatzel.
\newblock {Chemical Sensitivity of K$\beta$ and K$\alpha$ X-ray Emission from a Systematic Investigation of Iron Compounds}.
\newblock {\em Inorganic Chemistry}, 59(17):12518--12535, 2020.

\bibitem{Kowalska-2015}
Joanna Kowalska and Serena DeBeer.
\newblock The role of x-ray spectroscopy in understanding the geometric and electronic structure of nitrogenase.
\newblock {\em Biochimica et Biophysica Acta (BBA) - Molecular Cell Research}, 1853(6):1406--1415, 2015.

\bibitem{krishnan-2024}
Abiram Krishnan, Dong-Chan Lee, Ian Slagle, Sumaiyatul Ahsan, Samantha Mitra, Ethan Read, and Faisal~M. Alamgir.
\newblock Monitoring redox processes in lithium-ion batteries by laboratory-scale operando x-ray emission spectroscopy.
\newblock {\em ACS Applied Materials \& Interfaces}, 16(13):16096--16105, 2024.

\bibitem{giordanino-2014}
Filippo Giordanino, Elisa Borfecchia, Kirill~A. Lomachenko, Andrea Lazzarini, Giovanni Agostini, Erik Gallo, Alexander~V. Soldatov, Pablo Beato, Silvia Bordiga, and Carlo Lamberti.
\newblock Interaction of nh3 with cu-ssz-13 catalyst: A complementary ftir, xanes, and xes study.
\newblock {\em The Journal of Physical Chemistry Letters}, 5(9):1552--1559, 2014.

\bibitem{drager-1984}
G.~Dräger and O.~Brümmer.
\newblock {Polarized X‐Ray emission spectra of single crystals}.
\newblock {\em physica status solidi (b)}, 124(1):11--28, 1984.

\bibitem{czolbe-1992}
W.~Czolbe, U.~Dick, G.~Dräger, and K.~Fischer.
\newblock {The electronic structure of single crystal YBa2Cu3O7–$\delta$ studied by polarized Cu K$\beta$2,5 X-ray emission}.
\newblock {\em physica status solidi (b)}, 174(1):91--98, 1992.

\bibitem{bergmann-2002}
U.~Bergmann, J.~Bendix, P.~Glatzel, H.~B. Gray, and S.~P. Cramer.
\newblock {Anisotropic valence→core x-ray fluorescence from a [Rh(en)3][Mn(N)(CN)5]$\cdot$H2O single crystal: Experimental results and density functional calculations}.
\newblock {\em The Journal of Chemical Physics}, 116(5):2011--2015, 2002.

\bibitem{nishimoto-2012}
Satoshi Nishimoto, Stefan-Ludwig Drechsler, Roman Kuzian, Johannes Richter, Jiři Málek, Miriam Schmitt, Jeroen Van Den~Brink, and Helge Rosner.
\newblock {The strength of frustration and quantum fluctuations in LiVCuO4}.
\newblock {\em EPL (Europhysics Letters)}, 98(3):37007, 5 2012.

\bibitem{huang-2011}
Hsiao-Yu Huang, Nikolay~A. Bogdanov, Liudmila Siurakshina, Peter Fulde, Jeroen Van Den~Brink, and Liviu Hozoi.
\newblock {Ab initio calculation of d-d excitations in quasi-one-dimensional Cu d9 correlated materials}.
\newblock {\em Physical Review B}, 84(23), 12 2011.

\bibitem{roman-2023}
Marta Roman, Maria Fritthum, Berthold Stöger, Devashibhai~T. Adroja, and Herwig Michor.
\newblock {Charge density wave and crystalline electric field effects inTmNiC2}.
\newblock {\em Physical Review B}, 107(12), 3 2023.

\bibitem{maeda-2019}
Hiroyuki Maeda, Ryusuke Kondo, and Yoshio Nogami.
\newblock {Multiple charge density waves compete in ternary rare-earth nickel carbides, RNiC2(R:Y,Dy}.
\newblock {\em Physical Review B.}, 100(10), 9 2019.

\bibitem{kim-2013}
Jae~Nyeong Kim, Changhoon Lee, and Ji-Hoon Shim.
\newblock {Chemical and hydrostatic pressure effect on charge density waves of SmNiC2}.
\newblock {\em New Journal of Physics}, 15(12):123018, 12 2013.

\bibitem{drager-1976}
G.~Dräger and O.~Brümmer.
\newblock {X‐ray spectroscopic investigation of beryllium by polarized K‐emission valence bands}.
\newblock {\em physica status solidi (b)}, 78(2):729--735, 1976.

\bibitem{king-2014}
P.~D.~C. King, S.~McKeown Walker, A.~Tamai, A.~De~La~Torre, T.~Eknapakul, P.~Buaphet, S.-k. Mo, W.~Meevasana, M.~S. Bahramy, and F.~Baumberger.
\newblock {Quasiparticle dynamics and spin–orbital texture of the SrTiO3 two-dimensional electron gas}.
\newblock {\em Nature Communications}, 5(1), 2 2014.

\bibitem{wang-2023}
Yang Wang and Maciej Dendzik.
\newblock {Recent progress in angle-resolved photoemission spectroscopy}.
\newblock {\em Measurement Science and Technology}, 35(4):042002, 12 2023.

\bibitem{huang-2004}
D.~J. Huang, W.~B. Wu, G.~Y. Guo, H.~J. Lin, T.~Y. Hou, C.~F. Chang, C.~T. Chen, A.~Fujimori, T.~Kimura, H.~B. Huang, A.~Tanaka, and T.~Jo.
\newblock {Orbital ordering in La0.5 Sr1.5 Mn O4 studied by soft X-ray Linear Dichroism}.
\newblock {\em Physical Review Letters}, 92(8), 2 2004.

\bibitem{wu-2004}
W.B. Wu, D.J. Huang, G.Y. Guo, H.-j. Lin, T.Y. Hou, C.F. Chang, C.T. Chen, A.~Fujimori, T.~Kimura, H.B. Huang, A.~Tanaka, and T.~Jo.
\newblock {Orbital polarization of LaSrMnO4 studied by soft X-ray linear dichroism}.
\newblock {\em Journal of Electron Spectroscopy and Related Phenomena}, 137-140:641--645, 4 2004.

\bibitem{bergmann1999}
U~Bergmann, CR~Horne, TJ~Collins, JM~Workman, and SP~Cramer.
\newblock Chemical dependence of interatomic x-ray transition energies and intensities--a study of mn k$\beta$ and k$\beta$2, 5 spectra.
\newblock {\em Chemical physics letters}, 302(1-2):119--124, 1999.

\bibitem{de-groot-2008}
Frank De~Groot and Akio Kotani.
\newblock {\em Core Level Spectroscopy of Solids}.
\newblock CRC Press, 1st edition, 2008.

\bibitem{laihia-1998}
R.~Laihia, K.~Kokko, W.~Hergert, and J.~A. Leiro.
\newblock {K-emission spectra of Zn, ZnS, and ZnSe within dipole and quadrupole approximations}.
\newblock {\em Physical Review B, Condensed matter}, 58(3):1272--1278, 1998.

\bibitem{ogasawara-2004}
H.~Ogasawara, K.~Fukui, and M.~Matsubara.
\newblock {Polarization dependence of X-ray emission spectroscopy}.
\newblock {\em Journal of Electron Spectroscopy and Related Phenomena}, 136(1-2):161--166, 2004.

\bibitem{sakurai-2020}
J.~J. Sakurai and Jim Napolitano.
\newblock {\em Modern Quantum Mechanics}.
\newblock Cambridge University Press, 3rd edition, 2020.

\bibitem{griffiths-2017}
David~J. Griffiths.
\newblock {\em Introduction to Electrodynamics}.
\newblock Cambridge University Press, 4th edition, 2017.

\bibitem{slater-1960}
John~Clarke Slater.
\newblock {\em Quantum Theory of Atomic Structure}.
\newblock McGraw-Hill, 1st edition, 1960.

\bibitem{haverkort-2012}
M.~W. Haverkort, M.~Zwierzycki, and O.~K. Andersen.
\newblock {Multiplet ligand-field theory using Wannier orbitals}.
\newblock {\em Physical Review B}, 85(16):5113, 2012.

\bibitem{sajeev-2008}
Y.~Sajeev, M.~Sindelka, and N.~Moiseyev.
\newblock {Hund’s multiplicity rule: From atoms to quantum dots}.
\newblock {\em The Journal of Chemical Physics}, 128(6):1101, 2008.

\bibitem{watanabe-1966}
Hiroshi Watanabe.
\newblock {\em {Operator methods in ligand field theory}}.
\newblock Prentice-Hall, 1st edition, 1966.

\bibitem{jeitschko-1986}
W.~Jeitschko and M.H. Gerss.
\newblock {Ternary carbides of the rare earth and iron group metals with CeCoC2- and CeNiC2-type structure}.
\newblock {\em Journal of the Less Common Metals}, 116(1):147--157, 2 1986.

\bibitem{pacchioni-1987}
Gianfranco Pacchioni and Piercarlo Fantucci.
\newblock {Spin states and quenching of magnetism in naked and carbonylated nickel clusters}.
\newblock {\em Chemical Physics Letters}, 134(5):407--412, 1987.

\bibitem{jahrman-2019}
Evan~P. Jahrman, William~M. Holden, Alexander~S. Ditter, Devon~R. Mortensen, Gerald~T. Seidler, Timothy~T. Fister, Stosh~A. Kozimor, Louis F.~J. Piper, Jatinkumar Rana, Neil~C. Hyatt, and Martin~C. Stennett.
\newblock {An improved laboratory-based x-ray absorption fine structure and x-ray emission spectrometer for analytical applications in materials chemistry research}.
\newblock {\em Review of Scientific Instruments}, 90(2):024106, 2019.

\bibitem{chen-2025}
Yeu Chen, Anthony~J. Gironda, Yaxin Shen, André~D. Taylor, and Gerald~T. Seidler.
\newblock A ray tracing survey of asymmetric operation of the x-ray rowland circle using spherically bent crystal analyzers.
\newblock {\em J. Anal. At. Spectrom.}, 40:836--847, 2025.

\bibitem{bergmann-cramer-1998}
Uwe Bergmann and Stephen~P. Cramer.
\newblock {High-resolution large-acceptance analyzer for x-ray fluorescence and Raman spectroscopy}.
\newblock In Albert~T. Macrander, Andreas~K. Freund, Tetsuya Ishikawa, and Dennis~M. Mills, editors, {\em Crystal and Multilayer Optics}, volume 3448, pages 198 -- 209. International Society for Optics and Photonics, SPIE, 1998.

\bibitem{rovezzi-2017}
Mauro Rovezzi, Christophe Lapras, Alain Manceau, Pieter Glatzel, and Roberto Verbeni.
\newblock High energy-resolution x-ray spectroscopy at ultra-high dilution with spherically bent crystal analyzers of 0.5 m radius.
\newblock {\em Review of Scientific Instruments}, 88(1):013108, 01 2017.

\bibitem{hecht2016optics}
Eugene Hecht.
\newblock {\em Optics}.
\newblock Pearson, 5th edition, 2016.

\bibitem{grams-2019}
Christoph~P. Grams, Severin Kopatz, Daniel Brüning, Sebastian Biesenkamp, Petra Becker, Ladislav Bohatý, Thomas Lorenz, and Joachim Hemberger.
\newblock {Evidence for polarized nanoregions from the domain dynamics in multiferroic LiCuVO4}.
\newblock {\em Scientific Reports}, 9(1), 3 2019.

\bibitem{kas-2021}
J.~J. Kas, F.~D. Vila, C.~D. Pemmaraju, T.~S. Tan, and J.~J. Rehr.
\newblock {Advanced calculations of X-ray spectroscopies with FEFF10 and Corvus}.
\newblock {\em Journal of Synchrotron Radiation}, 28(6):1801--1810, 2021.

\bibitem{jahrman-2020}
Evan~P. Jahrman, William~M. Holden, Niranjan Govind, Joshua~J. Kas, Jatinkumar Rana, Louis F.~J. Piper, Carrie Siu, M.~Stanley Whittingham, Timothy~T. Fister, and Gerald~T. Seidler.
\newblock {Valence-to-core X-ray emission spectroscopy of vanadium oxide and lithiated vanadyl phosphate materials}.
\newblock {\em Journal of Materials Chemistry A}, 8(32):16332--16344, 2020.

\bibitem{mortensen-2017}
D.~R. Mortensen, G.~T. Seidler, Joshua~J. Kas, Niranjan Govind, Craig~P. Schwartz, Sri Pemmaraju, and David~G. Prendergast.
\newblock {Benchmark results and theoretical treatments for valence-to-core x-ray emission spectroscopy in transition metal compounds}.
\newblock {\em Physical Review B}, 96(12):125136, 2017.

\bibitem{abramson-2021}
Jared~E. Abramson, Nancy~M. Avalos, Agathe L.~M. Bourchy, Sarah~A. Saslow, and Gerald~T. Seidler.
\newblock {An exploration of benchtop X‐ray emission spectroscopy for precise characterization of the sulfur redox state in cementitious materials}.
\newblock {\em X-Ray Spectrometry}, 51(2):151--162, 2021.

\bibitem{dhakal-2023}
Diwash Dhakal, Darren~M. Driscoll, Niranjan Govind, Andrew~G. Stack, Nikhil Rampal, Gregory Schenter, Christopher~J. Mundy, Timothy~T. Fister, John~L. Fulton, Mahalingam Balasubramanian, and Gerald~T. Seidler.
\newblock {The evolution of solvation symmetry and composition in Zn halide aqueous solutions from dilute to extreme concentrations}.
\newblock {\em Physical Chemistry Chemical Physics}, 25(34):22650--22661, 2023.

\bibitem{miajaavila-2020}
Luis Miaja‐Avila, Galen~C. O'Neil, Young~Il Joe, Kelsey~M. Morgan, Joseph~W. Fowler, William~B. Doriese, Brianna Ganly, Deyu Lu, Bruce Ravel, Daniel~S. Swetz, and Joel~N. Ullom.
\newblock {Valence‐to‐core X‐ray emission spectroscopy of titanium compounds using energy dispersive detectors}.
\newblock {\em X-Ray Spectrometry}, 50(1):9--20, 2020.

\bibitem{gironda-2024}
Anthony~J. Gironda, Jared~E. Abramson, Yeu Chen, Mikhail Solovyev, George~E. Sterbinsky, and Gerald~T. Seidler.
\newblock {Asymmetric Rowland circle geometries for spherically bent crystal analyzers in laboratory and synchrotron applications}.
\newblock {\em Journal of Analytical Atomic Spectrometry}, 39(5):1375--1387, 2024.

\bibitem{tetef-2021}
Samantha Tetef, Niranjan Govind, and Gerald~T. Seidler.
\newblock {Unsupervised machine learning for unbiased chemical classification in X-ray absorption spectroscopy and X-ray emission spectroscopy}.
\newblock {\em Physical Chemistry Chemical Physics}, 23(41):23586--23601, 2021.

\end{thebibliography}

\end{document}